\documentclass[prb,twocolumn,preprintnumbers,amsmath,amssymb,floatfix,superscriptaddress]{revtex4}

\usepackage{graphicx}
\usepackage{color,soul}
\setstcolor{red}
\usepackage{multirow}
\usepackage[sort&compress]{natbib}
\usepackage{multirow}

\newcommand{\meV}{\ensuremath{\,\mbox{meV}}}

\newcommand{\mueV}{\ensuremath{\,\mu\mbox{eV}}}
\newcommand{\mT}{\ensuremath{\,\mbox{mT}}}
\newcommand{\Tesla}{\ensuremath{\,\mbox{T}}}
\newcommand{\nm}{\ensuremath{\,\mbox{nm}}}

\newcommand{\cm}{\ensuremath{\,\mbox{cm}}}

\newcommand{\vc}[1]{\ensuremath{\vec{#1}}}

\newcommand{\Hop}{\ensuremath{\mathcal{H}}}
\newcommand{\Hmb}{\ensuremath{\mathcal{H}}}

\newcommand{\sgn}{\ensuremath{\mbox{sgn}}}

\begin{document}

\title{Six-electron semiconductor double quantum dot qubits}
\author{Erik Nielsen}
\affiliation{Sandia National Laboratories, Albuquerque, New Mexico 87185, USA}
\author{Edwin Barnes}
\affiliation{Condensed Matter Theory Center, Department of Physics, University of Maryland, College Park, MD 20742, USA}
\author{J.~P.~Kestner}
\affiliation{Department of Physics, University of Maryland Baltimore County, Baltimore, MD 21250, USA}
\affiliation{Condensed Matter Theory Center, Department of Physics, University of Maryland, College Park, MD 20742, USA}
\author{S.~Das Sarma}
\affiliation{Condensed Matter Theory Center, Department of Physics, University of Maryland, College Park, MD 20742, USA}
\date{\today}

\begin{abstract}
We consider a double-quantum-dot (DQD) qubit which contains six electrons instead of the usual one or two.  In this spin qubit, quantum information is encoded in a low-lying singlet-triplet space much as in the case of a two-electron DQD qubit.  We find that initialization, manipulation, and read-out can be performed similarly to the two-electron case, and that energy gaps remain large enough that these operations can be performed robustly.  We consider DQD potentials with parameters chosen to be representative of current experimental capabilities.  Results are obtained using two complementary full configuration interaction methods.
\end{abstract}

\maketitle

The gate model for quantum computation requires the coupling of many individual quantum bits in order to perform quantum ``gates'' which in turn implement useful quantum algorithms.\cite{Childs_Algorithsm_RevModPhys_2010,Shor_1997,Nielsen_Chuang_QMbook}  There have been many proposed ways of realizing a single quantum bit (qubit), in a variety of material systems. Examples of experimentally demonstrated solid state qubits include the spin of a donor-bound electron\cite{Pla_SingleAtomQubit_2012}, the charge state of a single electron confined to a double-quantum-dot (DQD)\cite{Gorman_DQDChargeQubit_2005,Petta_chargeManipulationInDQD_2004}, and the total spin of two electrons confined to a DQD.\cite{PettaScience_2005,TaylorNatPhys_2005}  In this work, we consider an extension of this latter type: a qubit defined by the spin of \emph{six} electrons confined in a DQD. This system has recently been realized experimentally.\cite{Marcus_6eMMtalk_2012}

A qubit can be defined by the low energy singlet and unpolarized triplet states of two electrons confined in a double-quantum-dot (DQD) potential.\cite{PettaScience_2005}  The splitting between singlet and triplet states, called the \emph{exchange energy}, determines the rate of qubit rotation about the z-axis of the Bloch sphere and can be controlled quickly via electric gates by ``tilting'' the DQD, that is, raising the potential of one dot relative to the other.\cite{TaylorNatPhys_2005}  Qubit rotation about the x-axis of the Bloch sphere occurs when there is an effective magnetic field gradient across the two dots, either naturally occurring (as in GaAs) or applied by micro-magnets\cite{Obata_micromagnet_2010} or g-factor engineering\cite{Kato_gTensorMod_2003} (as would be needed in isotopically pure silicon). DQD qubits can be coupled electrostatically to form entangling gates equivalent to the standard CPHASE or CNOT gates up to single qubit rotations.\cite{Stepanenko_CoupledDQDs_2007}  Coherent single qubit rotations have been demonstrated in gallium arsenide (GaAs)\cite{PettaScience_2005} as well as silicon-germanium (SiGe)\cite{Maune_HRL_SiGeDQD_2012} in lateral DQDs, and an entangling gate between qubits has been demonstrated in GaAs.\cite{Shulman_Yacoby_CoupledDQDs_2012}

While the solid state environment has great promise for future scalability, semiconductor heterostructures are inherently ``dirty'', containing many defects and impurities even using state-of-the-art fabrication techniques.  Particularly problematic are charge impurities, which are present in concentrations of order $10^{10}\cm^{-2}$ in GaAs and $10^{11}\cm^{-2}$ in Silicon MOS.\cite{Nguyen_ImpurityEffectsInDQDs_2011} 
These impurities result in a potential landscape that is very rough when there are few conduction band electrons present, making it difficult to empty a quantum dot down to its last electron.  Additionally, if the impurities are not static, the resulting fluctuating charge can cause qubit decoherence.\cite{Witzel_longPRB_2012}  While singlet-triplet qubits are more robust to charge noise than charge qubits, to the extent that the singlet and triplet states defining the qubit space have differing charge distributions (e.g., during an exchange gate), charge fluctuations will still result in gate errors.

A natural way to reduce the impact of randomly placed charge impurities is to consider quantum dots containing more than one electron.  It has been shown that the exchange energy of double quantum dots containing more than one electron per dot is less sensitive to a charge impurity than that of a two-electron DQD.\cite{Barnes_Kestner_Screening_2011}  In this work, we demonstrate that a six-electron DQD forms a viable qubit which can be initialized, manipulated, and read out in a similar manner to a two-electron DQD qubit.  We show that in addition to the reduced sensitivity to charge impurities such many-electron qubits offer richer control capabilities than two-electron qubits. This furthermore obviates the need to deplete a dot of all but a single electron, a process that at best requires substantial time and resources, and at worst is exhausting to the point that the device is deemed unusable.  The choice of six electrons in a DQD is motivated by the fact that three electrons in a single dot result in a single valence electron (after filling the lowest S-type shell) so that behavior similar to a DQD with one electron per dot is expected.  (This argument assumes a single valley, an assumption we operate under throughout this work.)  Picturing a quantum dot as an artificial atom containing shells of electrons relies on the dot's orbital energy spacing being large compared to electron-electron correlation energy.  This regime is accessible to experiment, as evidenced by predictable shell-filling data in both GaAs\cite{Tarucha_spinFilling_verticalGaAs,Ciorga_spinFilling_lateralGaAs} and silicon dots\cite{Lim_spinFilling_lateralMOS_2011}, but is not necessarily the regime of all dots.  In particular, the dot confinement along with the electron effective mass set the orbital energy scale, which need not be larger than the energy scale of the Coulomb interaction, especially in the case of large dots.


We investigate the spectrum of an idealized DQD potential using two complementary configuration interaction (CI) techniques.  The CI method was first applied to multi-electron quantum dots within the quantum computing context over a decade ago,\cite{Xuedong_Multielectron_2001} with the qualitative conclusion that dots with an odd number of electrons should suffice as qubits.  We first consider an unrealistically small DQD for which both of our CI methods readily converge and which clearly illustrates the common qualitative behavior found for all the cases of interest.  Qualities of the six-electron spectrum related to the initialization, manipulation, and read-out are compared with those of a two-electron DQD which has the same valence electron tunneling rate, which we define below.  The use of this metric helps to isolate the multi-electron effects from those which occur by simply changing the number of electrons in the system.  We later calculate the six-electron spectra using parameters more compatible with today's experimentally realizable DQDs, and find that comparable qualitative behavior can be found with these larger-sized dots.  We use GaAs material parameters throughout the text. 


\section{System\label{secSystem}}

\begin{figure}[h]
\begin{center}
\includegraphics[width=2.2in,angle=270]{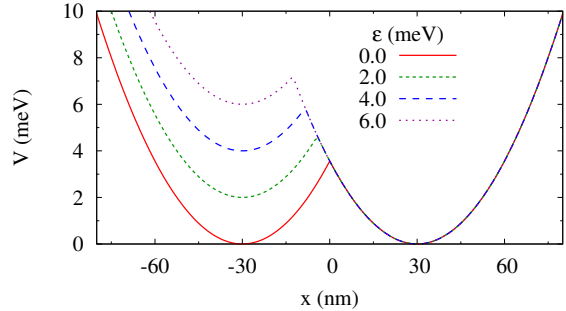}
\caption{DQD potential along x-axis, $V(x,0)$, given by Eq.~\ref{eqPot} with $L=30\nm$ and $E_0=3.0\meV$.  $V$ is the minimum of parabolas centered at $x = \pm L$ with curvature proportional to $E_0^2$. \label{figPotential}}
\end{center}
\end{figure}

The double quantum dot systems we consider are treated within the effective mass approximation with an electrostatic potential $V$ given as the minimum of two parabolic potentials,
\begin{equation}
\begin{array}{c}
V(x,y) = \frac{1}{2} m^*\omega^2 \min\left[ (x-L)^2 + \frac{\epsilon}{2}, (x+L)^2 - \frac{\epsilon}{2}\right] \\
\vspace{-0.1cm}\\
 + \frac{1}{2}m^* \omega^2 y^2 \,.
\end{array} \label{eqPot}
\end{equation}
There is assumed to be strong vertical ($z$-direction) confinement due to the semiconductor heterostructure, and we model only the lateral ($x$ and $y$) dimensions.  The potential of Eq.~\ref{eqPot} is parametrized by $\epsilon$, the energy difference or ``tilt'' between the left and right minima, $L$, half the distance between the minima, and $E_0 = \hbar\omega$, the confinement energy of the dots (in the lateral directions).  The dots we consider are isotropic, that is, the same confinement exists in both lateral directions.  We discuss this restriction below in the context of applying our results to realistic dots.  We have in mind $\epsilon$ as the primary control parameter, which is adjusted by varying the voltages applied to electrostatic gates.  Changing $\epsilon$ changes the exchange energy and thus the rate of the qubit rotation.  Plots of the potential along the $x$-axis for different $\epsilon$ are shown in Fig.~\ref{figPotential}.

Typical dots in GaAs have $E_0 = 1 - 10\meV$, $L = 25 - 100\nm$, and $\epsilon$ in the milli-electronVolt range. 

\section{Methods\label{secMethods}}

The configuration interaction algorithm\cite{CImethodBook} is well suited to modeling the multi-electron physics of small systems and has proven to be the method of choice when studying few electron systems in which multi-particle coherences are important\cite{Xuedong_Multielectron_2001,Nielsen_lowNoiseExchange_2010,Nguyen_ImpurityEffectsInDQDs_2011,Yang_LowNoiseCPhase_2011,NielsenRahman_ChargeDefects_2012}.  It consists of building antisymmetric multi-electron states from products of single electron states (Slater determinants) and diagonalizing the Hamiltonian in this multi-electron basis.  Since the dimension of the multi-electron state space grows rapidly with the number of electrons $N$ and the size of the single particle basis $M$ (roughly as $_MC_N$), even for a six-electron system the multi-electron basis must be ``truncated'', that is, reduced from it's full dimension.  The degree of accuracy achieved by a CI implementation is dependent on the single particle basis chosen (its size and form) and upon the truncation strategy and amount.  Configuration interaction techniques, even when truncated, are \emph{variational}, meaning that they will never underestimate the ground state energy (or the lowest energy state of a given symmetry class, e.g.~triplet states).  This property provides a means of comparing different CI implementations (see Appendix \ref{CIAppendix}) in the absence of a known exact solution, as the CI which produces the lower energy provides a stricter (better) bound on the ground state energy.

In this work, the CI algorithm is used to solve a $n$-electron effective mass Hamiltonian of the form
\begin{equation}
\Hmb = \sum_i^n \Hop_i + \sum_{i < j}\frac{e^2}{\kappa |r_i-r_j|} \,,
\label{eqMBHam}
\end{equation}
where $\kappa$ is an effective dielectric constant, and the single particle Hamiltonian for the $i^{\mathrm{th}}$ particle is given by
\begin{equation}
\Hop_i = \frac{(\vc{p}-e\vc{A})^2}{2m^*} + V(\vc{r})+ g^*\mu_B\vc{B} \cdot \vc{S} \,.
\label{eq1PHam}
\end{equation}
We have used $\vc{r}$ and $\vc{p}$ to denote the position and momentum, respectively, of the $i^{\mathrm{th}}$ electron and $V$ is the single-particle potential (Eq.~\ref{eqPot}).  The effective mass $m^*$ is generally a tensor but is assumed to be a scalar here, as it is in GaAs and even silicon when dealing with only the lateral directions.  The magnetic field $\vc{B} = \vc{\nabla} \times \vc{A}$ is determined by the  vector potential $\vec{A}$.  We take $\vc{A}$ in the symmetric gauge, $\vc{A} = B/2(-y\hat{x}+x\hat{y})$.  Coupling between spin and magnetic field is given by the effective g-factor $g^*$ multiplied by the Bohr magneton $\mu_B$.  GaAs material parameters, $m^* = 0.067\,m_e$, $g^*=-0.44$, and $\kappa = 12.9$ are taken. 

We use two complementary CI implementations to study the six- and two-electron DQDs of this paper.  The first, which we call CI-1, uses the Fock-Darwin states as a single particle basis.  The Fock-Darwin states are the exact single-electron eigenstates of a 2D parabolic dot in the presence of a constant magnetic field, and have the form
\begin{eqnarray}
&&\phi^{+-}_{nm}(x,y) = \frac{1}{l_0}\sqrt{\frac{ \left(\frac{n-|m|}{2}\right)! }{ \pi\left(\frac{n+|m|}{2}\right)!}} \left( \frac{ x \pm x_0 + i y \,\sgn\, m }{l_0} \right)^{|m|} \nonumber \\
&& \times e^{-\frac{(x \pm x_0)^2 + y^2}{2l_0^2} \pm i\frac{x_0 y}{2l_B^2}} L_{\frac{n-|m|}{2}}^{|m|} \left( \frac{ (x \pm x_0)^2 + y^2}{ l_0^2 } \right),\label{eqFDstates}
\end{eqnarray}
where $l_0 = l_B/(1/4 + \omega^2/\omega_c^2)^{1/4}$, $l_B = \sqrt{\hbar c/ e B}$, $\omega_c = eB/m^*c$, $L_n^m(x)$ is an associated Laguerre polynomial, and the associated single particle energies are

\begin{equation}
E_{n,m} = (n+1)\sqrt{\frac{1}{4} + \frac{\omega^2}{\omega_c^2}} \hbar\omega_c + \frac{m}{2}\hbar\omega_c \,. \label{eqFDenergies}
\end{equation}

The second CI implementation, which we denote CI-2, uses a set of s-type Gaussian orbitals as a single particle basis.  We denote the number of spatial Gaussian functions $n_G$, which is half the number of elements in the single particle basis due to spin.  Each spatial part of a Gaussian orbital is centered at some point $(x_0,y_0)$ and made gauge invariant by placing a magnetic-field dependent factor in its exponent, with the resulting form
\begin{equation}
\begin{array}{c}
g(x,y) = \mathcal{N} e^{-\alpha(x-x_0)^2}e^{-\alpha(y-y_0)^2} \\
\hspace{2cm} \times\,e^{\frac{ieB}{2\hbar}\left(y_0x-x_0y\right)}\,,
\end{array}
\end{equation}
where $\mathcal{N}$ is a normalization factor and $B$ is the magnetic field.  The vectors $\vec{r}_0 = (x_0,y_0)$ and $\alpha$ set the position and size of the Gaussian function, respectively.  The sizes of the Gaussians are set to match that of the ground state of a single dot with the confinement energy $E_0$, so $\alpha = (m^*E_0)/2\hbar^2$.  The positions are set on a regular rectangular grid with spacing equal to one half of the Gaussian ``radius'' $1/\sqrt{\alpha}$ (a somewhat arbitrary value found to give good results in practice).

\begin{figure}
\begin{center}
\includegraphics[width=2.4in]{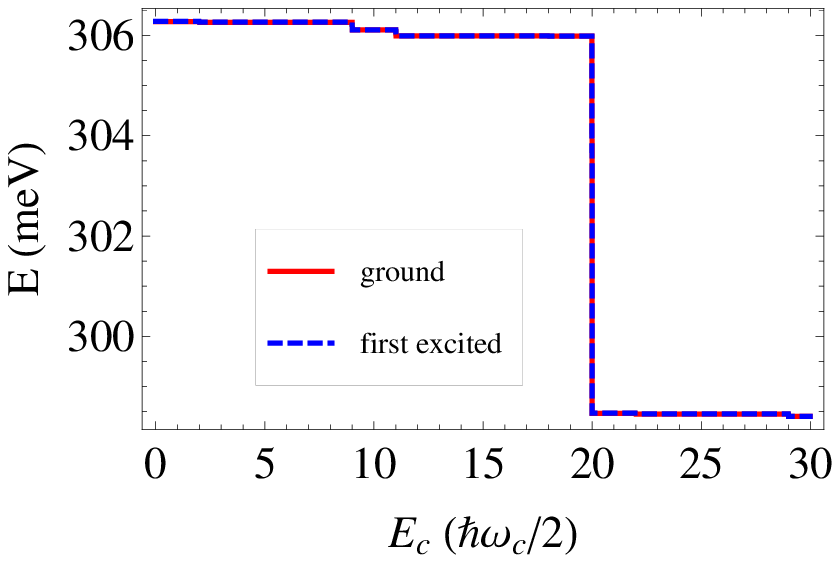}
\includegraphics[width=2in,angle=270]{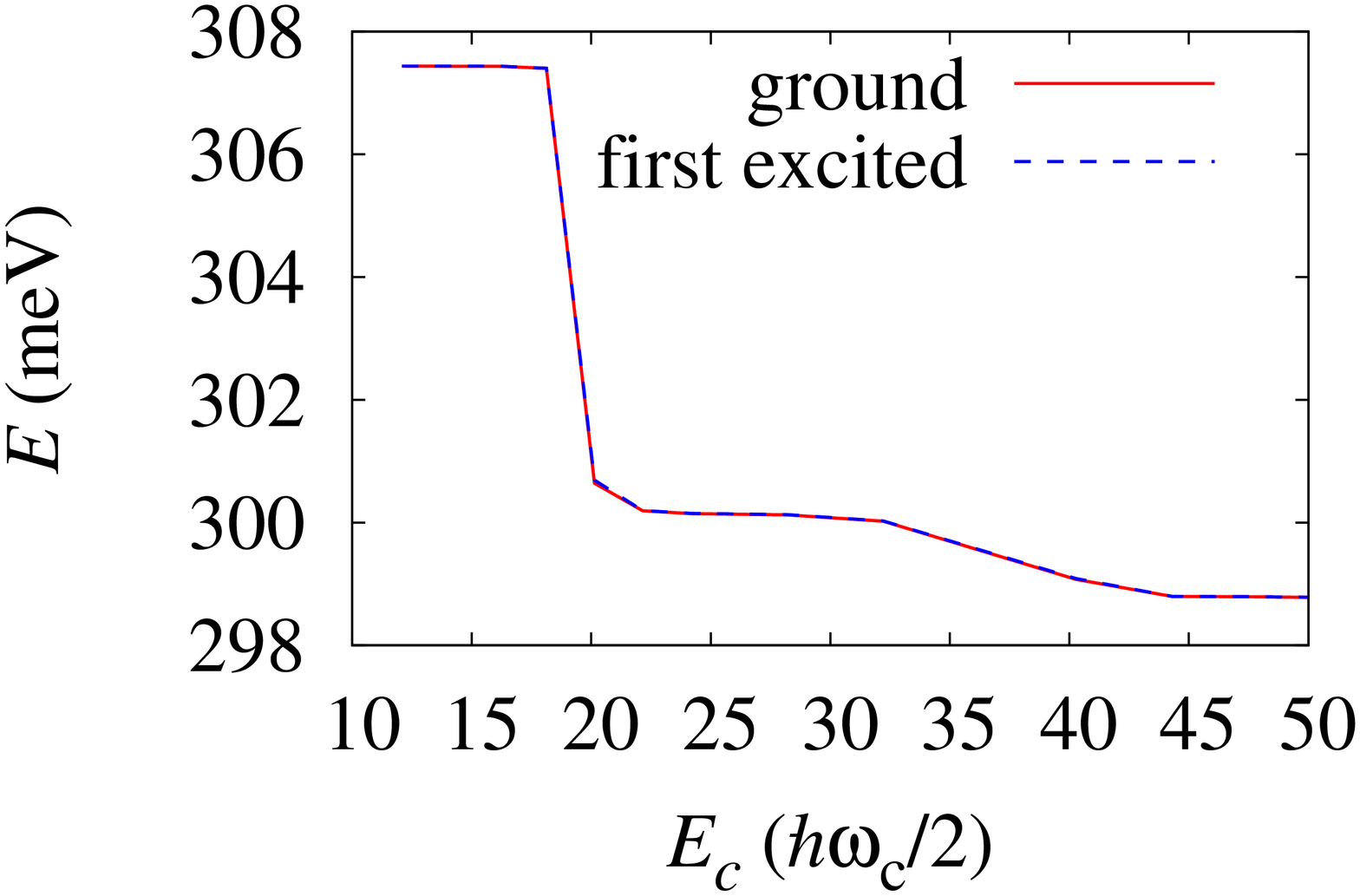}
\caption{Convergence of the lowest two $S_z=0$ levels as a function of the cutoff energy using CI-1 (upper) and CI-2 (lower).  In the lower plot, CI-2 uses a basis of $n_G=26$ Gaussian functions, arranged on each dot in a 3 by 3 array with one additional element along each compass direction.\label{figConvergenceSmall}}
\end{center}
\end{figure}

Apart from the basis being used, the two methods are identical (though they were developed separately and are completely independent).  Both truncate the full space of antisymmetrized product functions by keeping only those whose non-interacting energy (found by evaluating only the first term of Eq.~\ref{eqMBHam}) is at most a specified cutoff energy, $E_c$, above the non-interacting energy of the ground state.  We test each implementation for convergence by checking that the resulting multi-electron energies are converged with respect to $E_c$.  This convergence is specific to the set of DQD parameters, and is faster for smaller dots since greater confinement energy means that the low-energy many-electron states are almost entirely spanned by products of the lower orbital states kept by the CI.  Figure \ref{figConvergenceSmall} shows the convergence of the lowest single level for each method for the parameters $E_0=24.26\meV$, $L=20\nm$, $\epsilon=0$, and $B=2.8\Tesla$, which we motivate in section \ref{secResults} below.  Both CI methods converge to nearly the same energy, with the convergence of CI-1 showing an abrupt drop at $E_c=20\hbar\omega_c/2$.  This drop is due to the inclusion of the first excited orbital (Fock-Darwin state) with the same magnetic quantum number as each dot's valence orbital.  The relatively large Coulomb matrix elements between these orbitals results in the observed reduction in ground state energy.  The convergence of CI-2 is more gradual since Gaussians are not the excited eigenstates of a dot, but we see a similar decrease in the ground state energy near $E_c=20\hbar\omega_c/2$ which indicates the Gaussian basis is doing a decent job of spanning the excited single-dot levels.  We find that the drop in energy near $20\hbar\omega_c/2$ is larger in the case of CI-1, and that it takes a much larger energy cutoff and a large number of Gaussians (26 in Fig.~\ref{figConvergenceSmall}) to obtain as low a ground state energy using CI-2.  If CI-2 is used with fewer Gaussians a drop near $20\hbar\omega_c/2$ is still seen, but is smaller than for cases with a larger Gaussian basis. 


Since CI-1 uses the exact eigenstates of a single dot as its basis, it performs best when the confinement energy of dots are large (small dots spatially) so that there is a large gap between the valence orbital and those beneath it.  This results in multi-electron wave functions which have a character similar to the valence electron orbitals, as in the case of atomic physics.  The second CI implementation (CI-2) performs comparatively less well at tight confinement, measured by the number of configurations that need to be kept to achieve similar ground state energies.  The Gaussian basis of CI-2, however, is better able to produce multi-electron wave functions that are much larger (spatially) than the single-electron solutions of the same potential.  We use CI-1 to model cases where the confinement energy ($E_0$) is large, and CI-2 in cases where $E_0$ is small.  What constitutes a ``large'' or ``small'' $E_0$ is determined by comparing the resulting energies and convergence properties of both CI implementations for that particular case.  Both CI implementations were necessary for the results we present in this paper, and for each set of parameters below we indicate which implementation was used.

\section{6-electron DQD Qubit Spectra\label{secResults}}

%

The configuration interaction method is applied to several DQD and single-dot systems to assess the viability of a 6-electron DQD qubit.  In this section, we first explain and motivate the parameters that are used and then present the results.

\subsection{Parameter Choices}
A double quantum dot in our model is parameterized by a confinement energy $E_0$, an inter-dot separation $2L$, the magnetic field $B$, and a detuning or tilt $\epsilon$.  We have in mind a DQD qubit where $\epsilon$ is tuned by varying the voltages of electrostatic gates to control qubit rotation while the other parameters are held fixed, and thus the majority of the figures below plot energies vs.~$\epsilon$.  We consider three different sized 6-e DQDs, given by $E_0=24.26$, $7.28$, and $1.82\meV$.  These values of $E_0$ correspond to dot diameters of $13$, $25$, and $50\nm$, respectively, where the diameter $d$ is taken from the size of the harmonic oscillator ground state $d = 2\sqrt{ \hbar^2 / (m E_0) }$.  While the diameter is a good indicator of the size of a single electron dot, the multi-electron dots we study here may have larger spatial extent, and it is meant only to give a rough idea of the size of the confining potential.  Experimental DQDs more typically confine electrons into puddles of around $25-50\nm$ across, and so we understand the larger two dot sizes to be more realistic than the smallest size.   In many cases we lock the vertical magnetic field $B$ to $E_0$  by setting $\omega_c = 2\omega/\sqrt{99}$.  This relationship is chosen out of convenience for the numerical methods, to simultaneously avoid near-degeneracy effects (which would occur at too small a magnetic field) and the overlap of different shells in energy (which would occur at large magnetic field).\cite{Barnes_Kestner_Screening_2011}

\begin{figure}[h]
\begin{center}
\includegraphics[width=1.8in,angle=270]{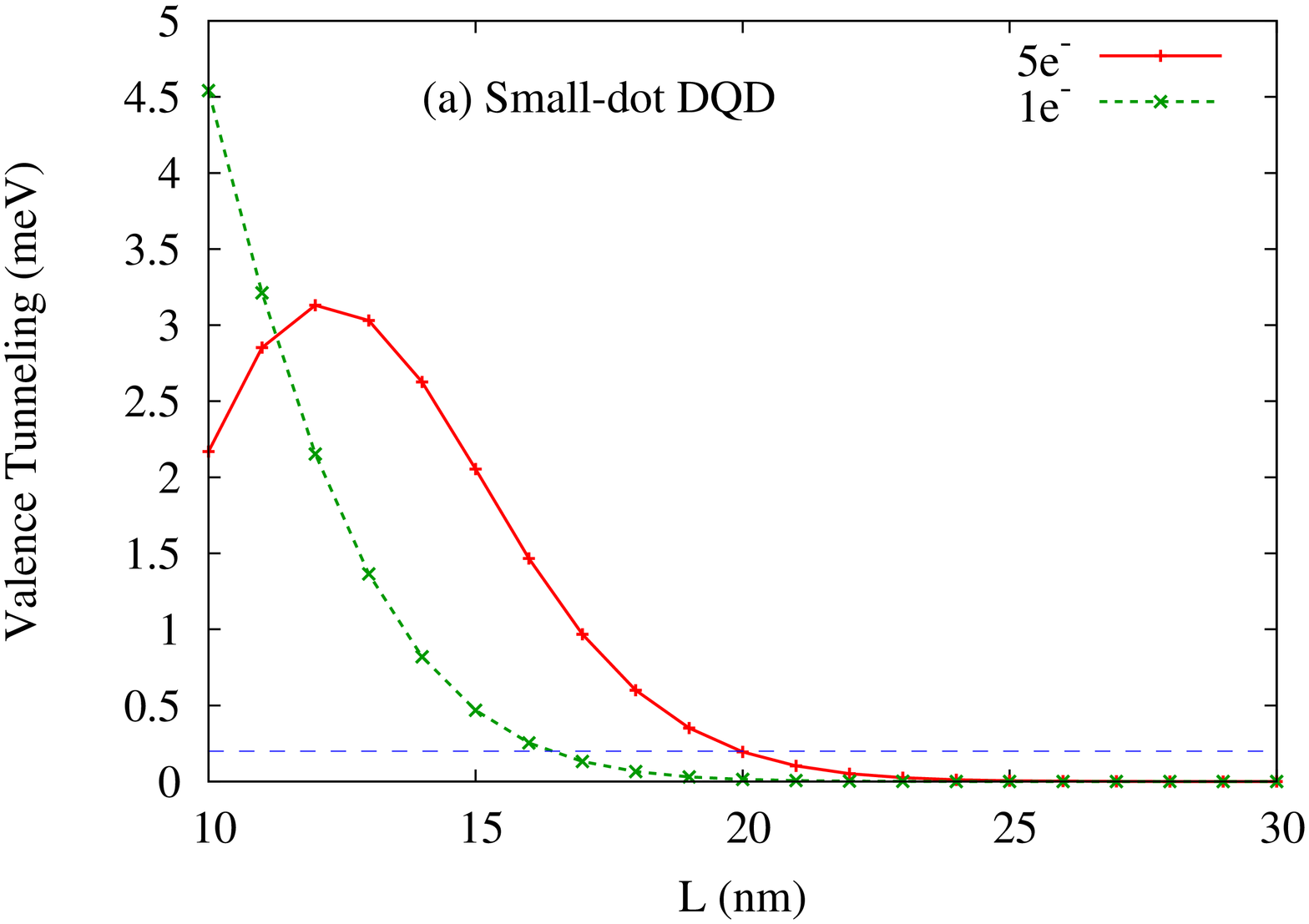}
\includegraphics[width=1.8in,angle=270]{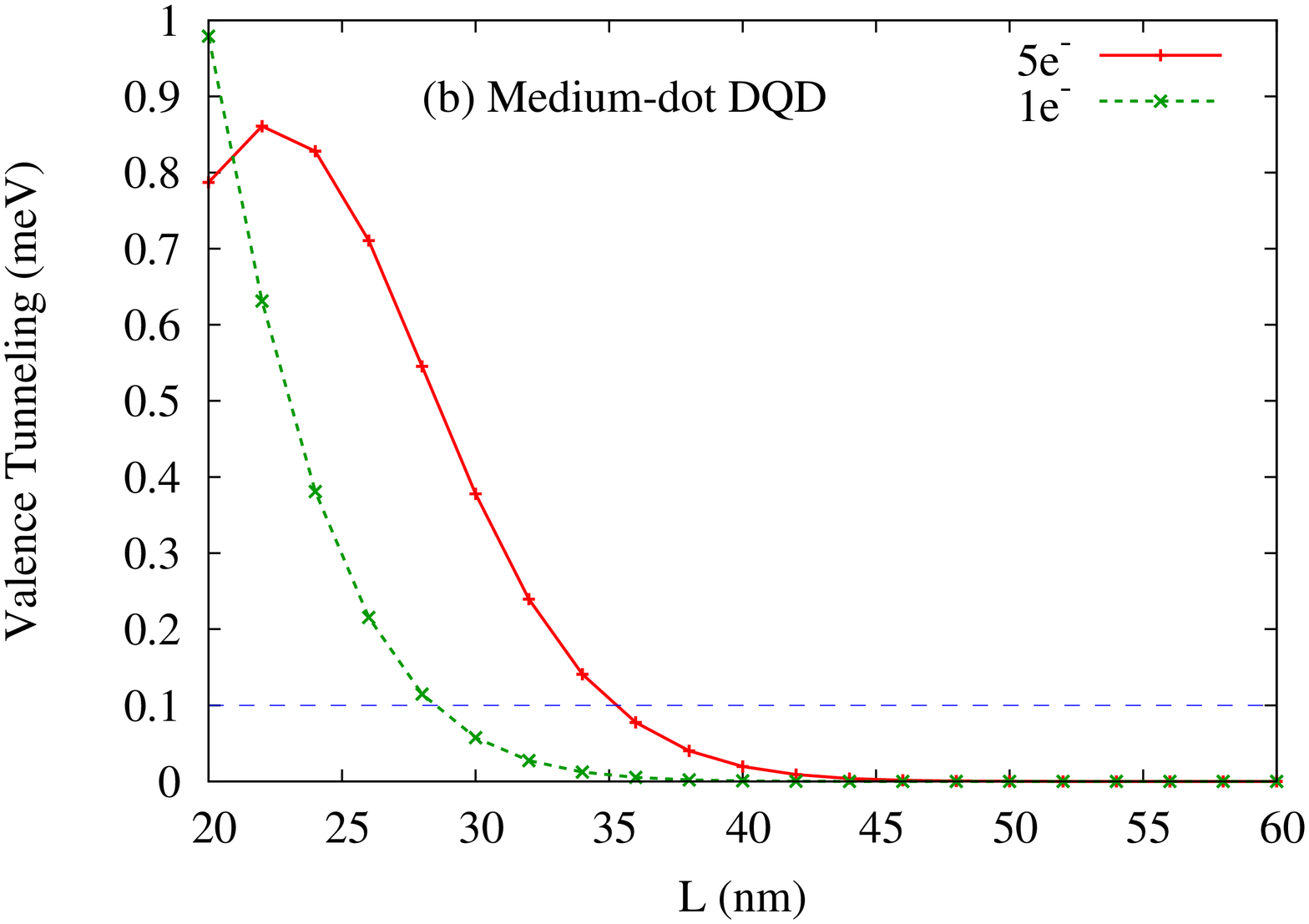}
\includegraphics[width=1.8in,angle=270]{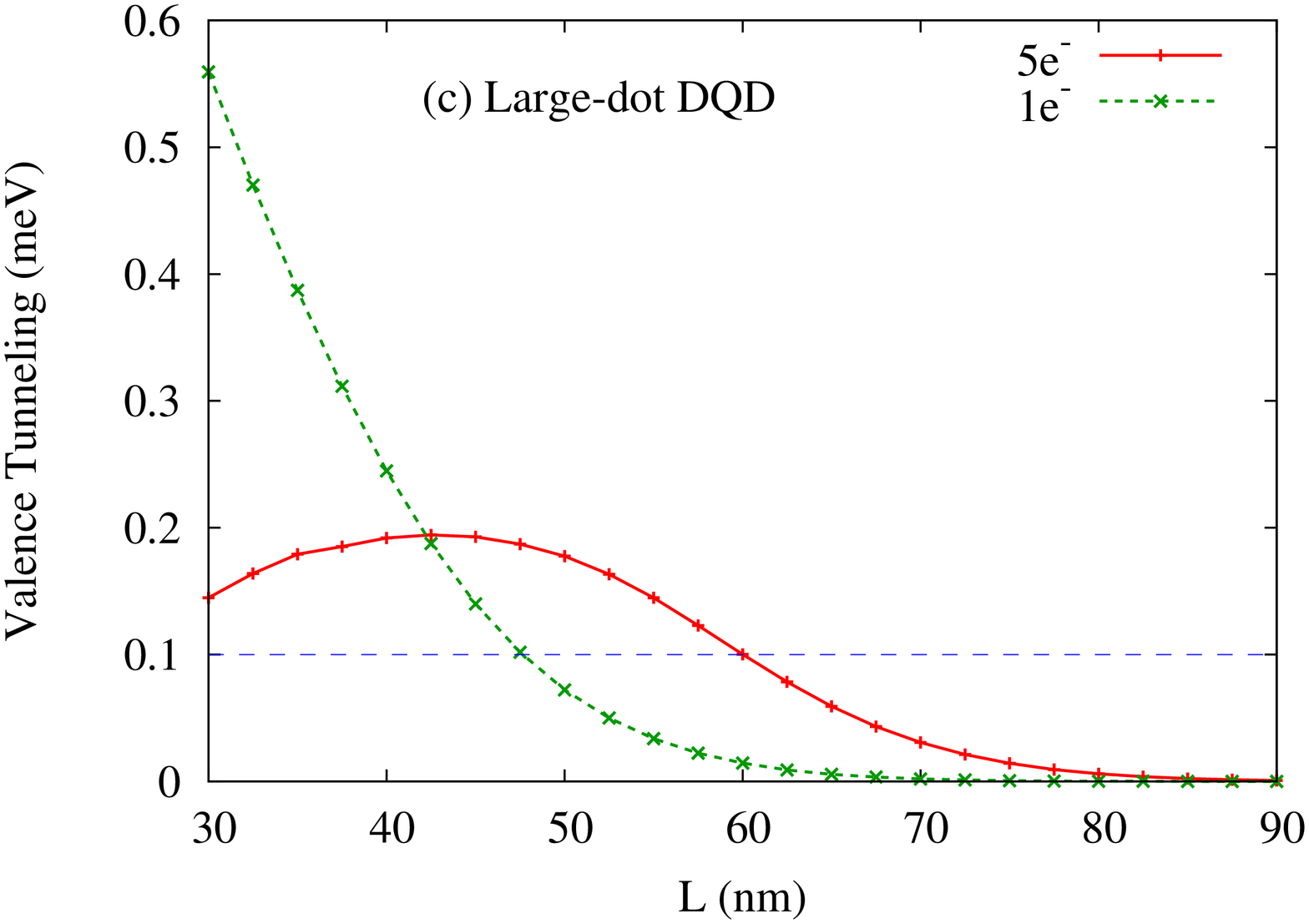}
\caption{Valence tunneling as a function of half the inter-dot separation $L$, for $E_0=24.26$, $7.28$, and $1.82\meV$ (shown in (a), (b), and (c), respectively).  A vertical magnetic field $B$ is set such such that $\hbar\omega_c = 2E_0/\sqrt{99}$ ($\omega_c=eB/m$). Horizontal lines indicate the constant value of the valence tunneling chosen to define the $L$ parameters in each case. Calculations were performed using CI-2 with $n_G=18$ and $E_c=55\meV$.\label{figValenceTunneling}}
\end{center}
\end{figure}

We choose the inter-dot separation to keep the dots separated enough so that charge sectors remain distinct yet close enough so that there is appreciable tunneling between the dots.  
Specifically, we define the ``valence tunneling'' of a DQD with $N$ electrons to be the energy difference between the ground and first excited state of the same system with $N-1$ electrons.  This estimates the energy required to move an electron from the valence level of one dot to the valence level of the other, and thus gives a metric of the tunneling between the dots.  The valence tunneling falls off with increasing inter-dot separation at large $L$, and we choose $L$ for each of the $E_0$ values above by somewhat arbitrarily choosing a value of the valence tunneling which lies on the tail of the valence tunneling vs.~$L$ curve.  For $E_0=24.26$, $7.28$, and $1.82\meV$ we choose values $0.2$, $0.1$, and $0.1$ for the valence tunneling, which result in values of $L$ equal to $20$, $35$ and $60\nm$, respectively (see Fig.~\ref{figValenceTunneling}).  (The decrease of valence tunneling as $L$ is reduced smaller than the dot diameter is spurious and signals the breakdown of the CI method.)

When we compare 6-electron DQD results with 2-electron DQD results, our 2-electron calculations use the same values of $E_0$ and $B$ but vary $L$ so that the valence tunneling remains the same as in the 6-electron calculation.  Essentially, the DQD potentials used for our 2-electron calculations have more closely spaced dots to maintain the same wave function overlap as in the 6-electron case (since the 2-electron wave functions are smaller than the 6-electron wave functions).  The reason we fix the valence tunneling instead of using the same potential for 2 and 6 electrons is to mimic the situation in a real device:  we suppose that when a realistic DQD is tuned between 2 and 6 electrons that the inter-dot tunneling is also tuned so as to obtain similar behavior of the DQD (e.g., charge stability lines and honeycomb charge sector structure).  This results in values of $L$ approximately equal to $16.5$, $28.5$, and $47.5\nm$ for the three values of $E_0$ above (see Fig.\ref{figValenceTunneling}).

 In summary, the DQD parameters just discussed  are given in Table \ref{tblParameters}, which we refer to as our ``small-dot'', ``medium-dot'', and ``large-dot'' DQD parameters as indicated.  We use these sets of parameters, and variations around them, for all of the results to follow.

\begin{table}[h]
\begin{tabular}{|c|c|c|c|c|}
\hline
 \multicolumn{5}{|c|}{ DQD Parameters } \\
\hline
 Name & $E_0$ & $L$ (6$e^-$) & $L$ (2$e^-$) & $B$ \\
\hline
Small-dot  & $24.26\meV$ & $20\nm$ & $16.5\nm$ & $2.8\Tesla$ \\
Medium-dot & $7.28\meV$  & $35\nm$ & $28.5\nm$ & $847\mT$ \\
Large-dot  & $1.82\meV$  & $60\nm$ & $47.5\nm$ & $212\mT$ \\
\hline
\end{tabular}
\caption{Summary of double quantum dot parameters.\label{tblParameters}}
\end{table}

In the small-dot DQD the orbital and multi-electron state energy spacings are large enough to result in a clear and relatively simple spectrum which we present first.   Then, since the small-dot parameters are unrealistic for actual devices, we focus on the medium- and large-dot DQD systems and consider how the many-electron spectrum evolves as one moves into a parameter regime which more closely reflects real devices.


\subsection{Small-Dot DQD Spectrum}

\begin{figure}[h]
\begin{center}
\includegraphics[width=2.8in]{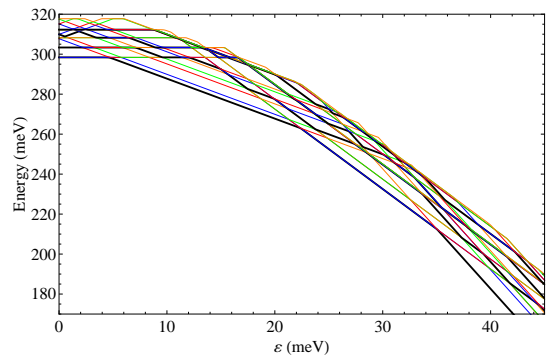}
\caption{6 electron energies vs. detuning using CI-1 for small-dot parameters ($E_0=24.26\meV$, $L=20\nm$, $B=2.8\Tesla$) and energy cutoff $E_c=30\hbar\omega_c/2=73.15\meV$. \label{fig6eSpectrumSmall}}
\end{center}
\end{figure}

The energy spectrum of a 6-electron DQD using our small-dot parameters (see Table \ref{tblParameters}) is shown in Fig.~\ref{fig6eSpectrumSmall}.  Only un-polarized, i.e., $S_z=0$ states are shown, and we arbitrarily cut off the upper part of the spectrum.  
We say a state is in ``charge sector'' $(n,m)$ when there are $n$ electrons in the left dot and $m$ electrons in the right dot.  The four slopes seen in Fig.~\ref{fig6eSpectrumSmall} correspond to the four possible charge sectors: (3,3), (2,4), (1,5), and (0,6).  (There are always equal or more electrons in the right dot since positive tilt ($\epsilon$) corresponds to the right dot having lower energy.)  We denote the $S_z=0$ energy levels by a capital S, T, or Q, indicating a spin-0 singlet, a spin-1 triplet state, or a spin-2 quintuplet state, followed by a zero-based index indicating the energy ordering.  For example, the lowest lying singlet and triplet states are denoted S0 and T0, respectively.  The progression of the lowest singlet eigenstate (S0) among the charge sectors is noted in Fig.~\ref{fig6eSpectrumSmall}.  The relative ``size'' of the regions where S0 is in each charge sector is an indication of the shell structure.  For example, the tilt required to remove an electron from a 3-electron dot and add it to another 3-electron dot is less then the tilt required to remove an electron from a 2-electron dot and add it to a 4-electron dot.

\begin{figure}[h]
\begin{center}
\includegraphics[width=1.5in]{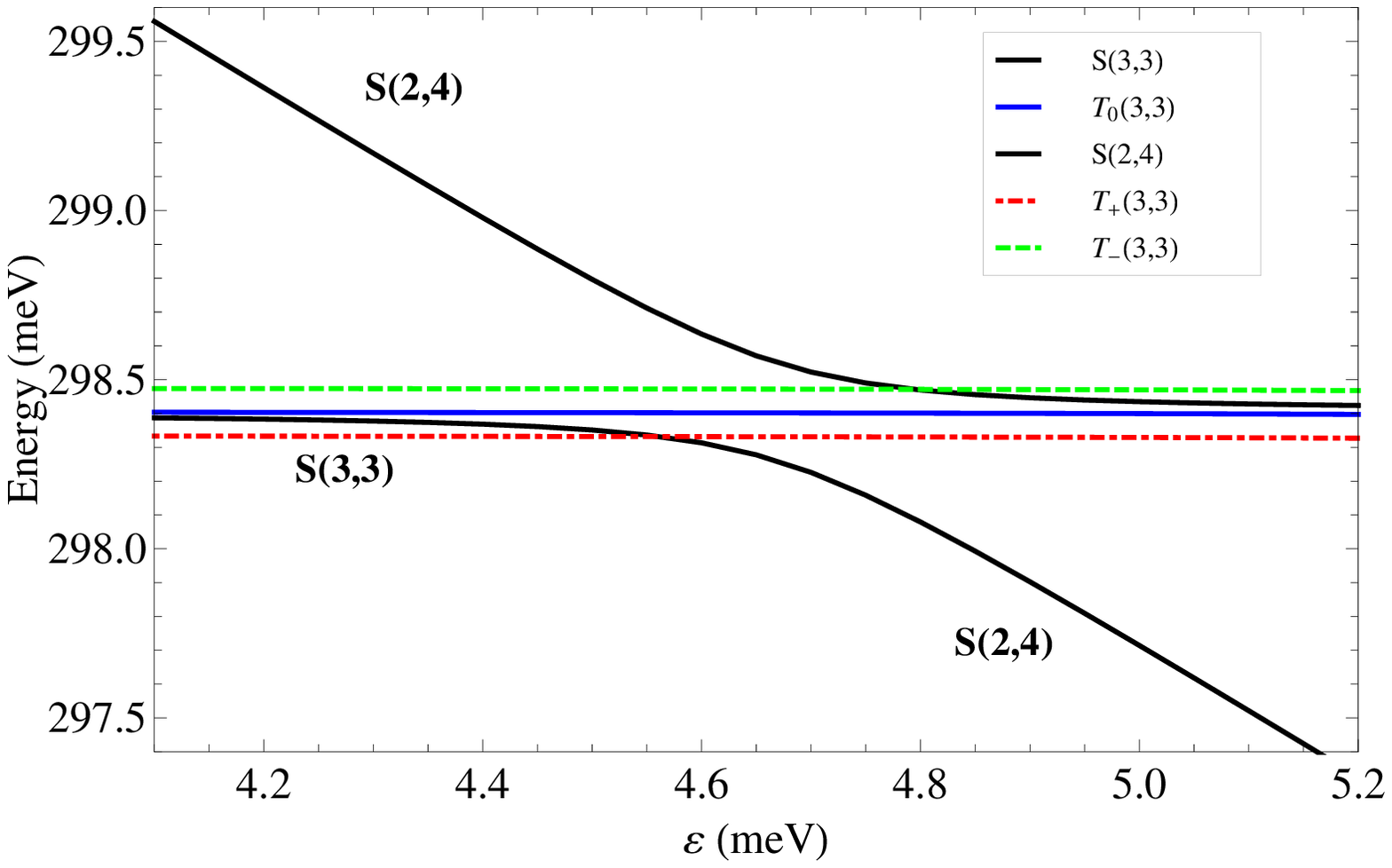} 
\includegraphics[width=1.5in]{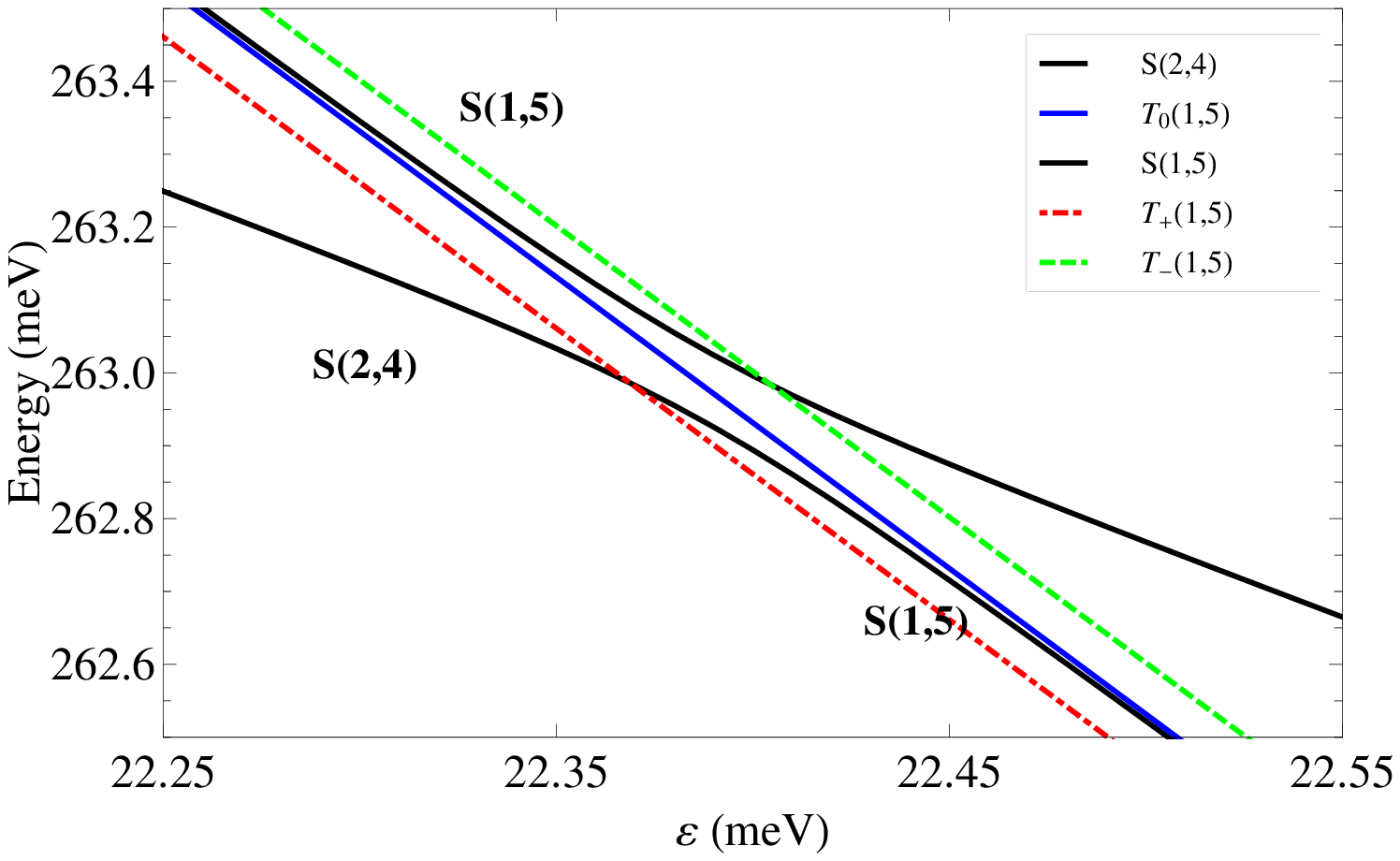} 
\includegraphics[width=1.5in]{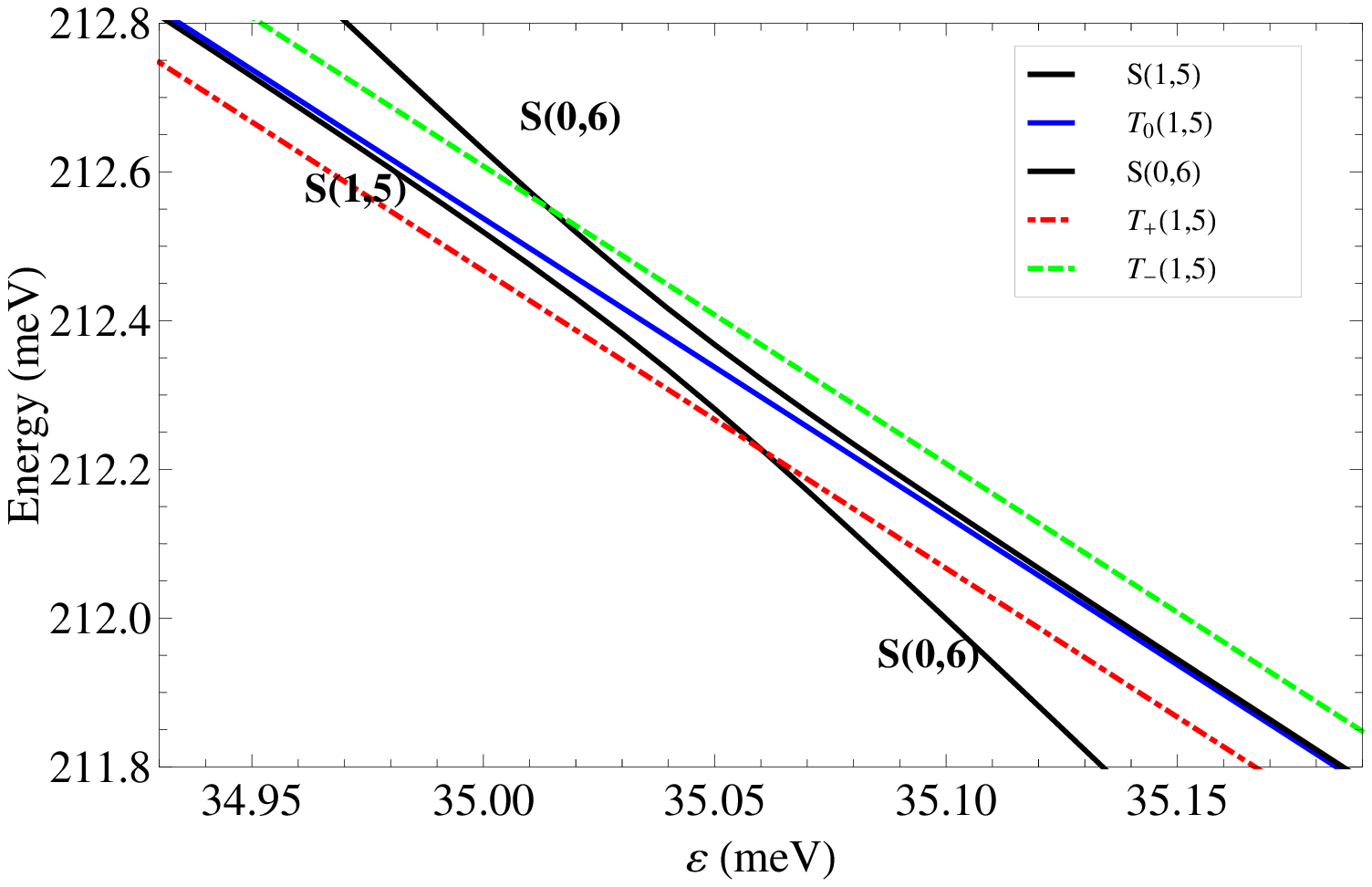} 
\caption{6 electron energies vs detuning using CI-1 for small-dot parameters and energy cutoff $E_c=30\hbar\omega_c/2=73.15\meV$, zoomed to each of the three anti-crossings, (3,3)-(2,4), (2,4)-(1,5), and (1,5)-(0,6) of the lowest singlet state, ``S0''.\label{fig6eSpectrumSmallZooms}}
\end{center}
\end{figure}

\begin{figure}[h]
\begin{center}
\includegraphics[width=1.5in]{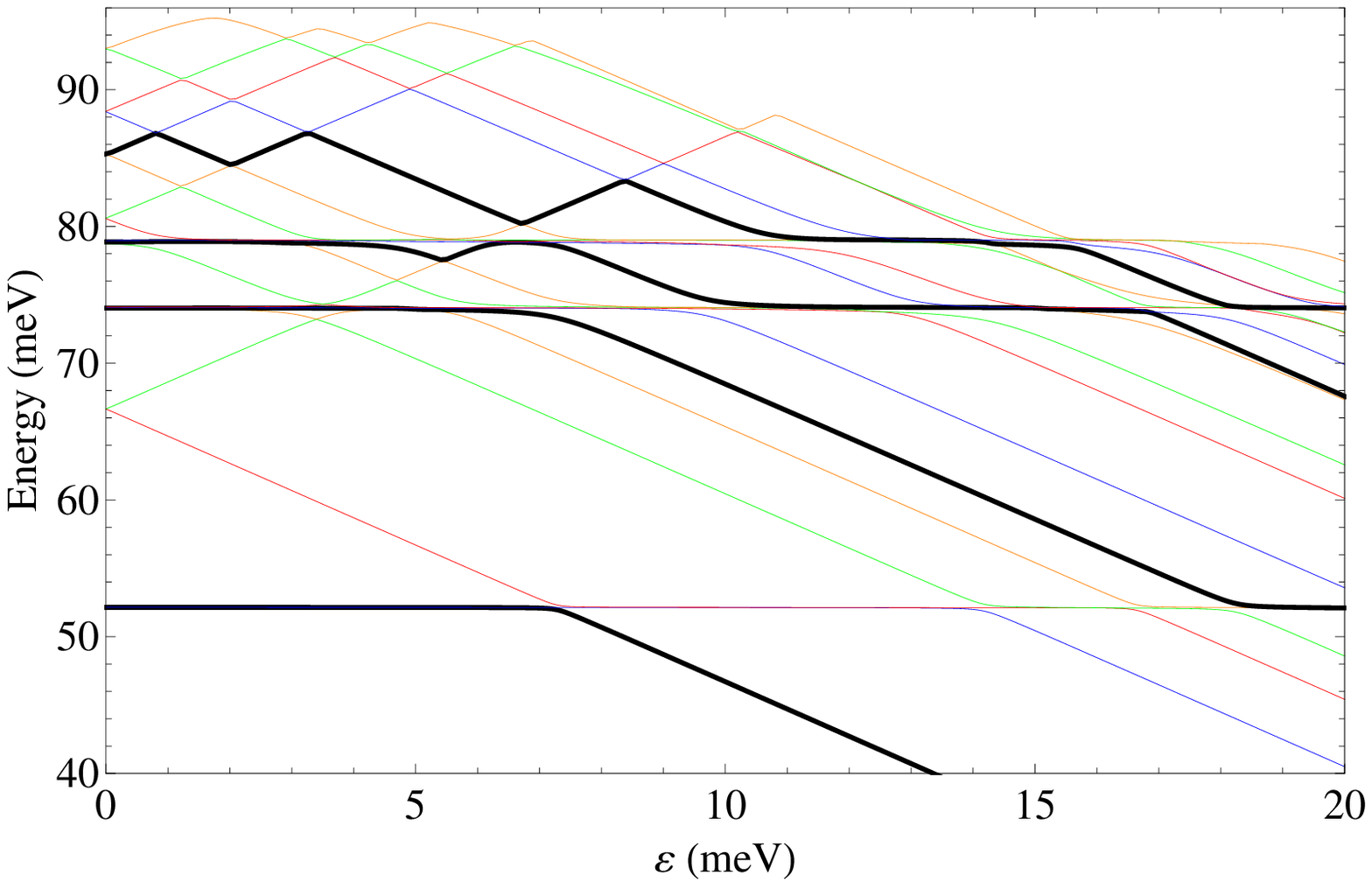}
\includegraphics[width=1.5in]{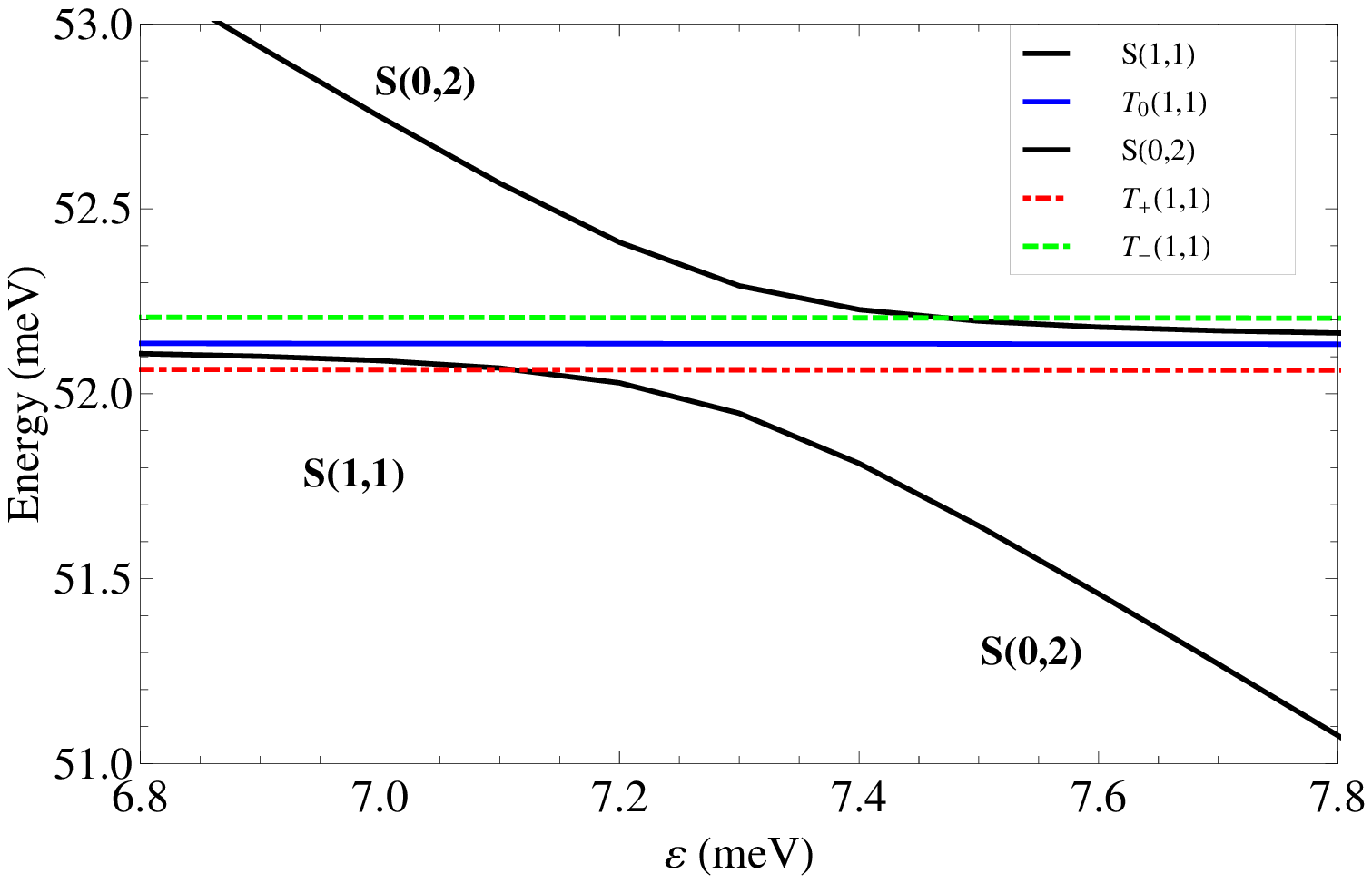}
\caption{2 electron energies vs detuning using CI-1 for small-dot parameters ($E_0=24.26\meV$, $L=16.5\nm$, $B=2.8\Tesla$) and energy cutoff $E_c=30\hbar\omega_c/2=73.15\meV$.  The left panel shows many of the 2-electron energies over a large range of detuning; the right plot shows a zoomed view of the lowest singlet anti-crossing.  S(1,1) and S(0,2) label the lowest singlet state in the (1,1) and (0,2) charge sectors, respectively, and (T$_-$,T$_0$,T$_+$) label the three lowest (1,1) triplet states with $S_z=-1$, $0$, and $1$, respectively. \label{fig2eSpectrumSmall}}
\end{center}
\end{figure}

We find that for our small-dot parameters the lowest two $S_z=0$ states are a singlet and triplet state, and that there is a clear order-meV gap to higher states.  When the singlet state of one charge sector crosses the singlet state of another there is an anti-crossing with a finite gap as shown by Fig.~\ref{fig6eSpectrumSmallZooms}, which zooms in on each of the singlet anti-crossings.  Each anti-crossing is qualitatively similar to the (1,1)-(0,2) crossing in the corresponding 2-electron spectrum shown in Fig.~\ref{fig2eSpectrumSmall}. Thus, the most basic criteria for a qubit are satisfied: there exists an isolated singlet-triplet ground space with variable splitting throughout a wide range of tuning.  We will discuss qubit characterization further below, but first let us consider how the energy spectrum changes with respect to the dot size.

\subsection{Medium- and Large-dot DQD Spectra}
Having seen that the 6-electron DQD can be operated as a qubit in the usual way in the case of small dots, we now turn to the question of whether this is also true of DQDs with larger dots.  This amounts to a question of energy scales, since by enlarging the dots we are decreasing the energy spacing between the orbitals of a single dot and consequently the spacing between the many-electron levels.  Ideally the spectrum would have an isolated 2-level $S_z=0$ ground space formed by a singlet and triplet, or by a singlet and the unpolarized state of a higher-spin multiplet, with the singlet state at lower energy.  The reason we desire a singlet ground state is for qubit initialization:  in order to initialize a $S_z=0$ state there must be a regime accessible to the qubit in which a singlet state is the true ground state with respect to all $S_z$ sectors. If a higher-spin state is the ground state, e.g.~a triplet state, then initialization is made more challenging if not impossible because it will always be energetically preferable to initialize into one of the polarized triplet states for any non-zero magnetic field along $z$.  Thus, for qubit operation analogous to the 2-electron DQD case, we require a singlet ground state in addition to the existence of an isolated 2-level manifold with a tunable energy splitting.

\begin{figure}[h]
\begin{center}
\includegraphics[width=2in,angle=270]{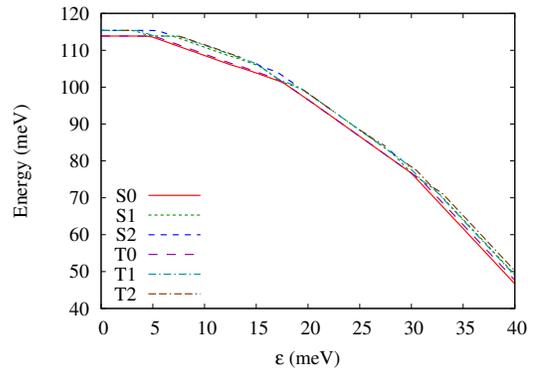}
\caption{6 electron energies vs detuning using CI-2 ($n_G=18$, $E_c=55\meV$) for medium-dot parameters ($E_0=7.28\meV$, $L=50\nm$, $B=847\mT$). $S_z=0$ for all states, and labels denote spin (S=singlet, T=triplet) and energy-ordering index. \label{fig6eSpectrumMedium}}
\end{center}
\end{figure}

\begin{figure}[h]
\begin{center}
\includegraphics[width=1in,angle=270]{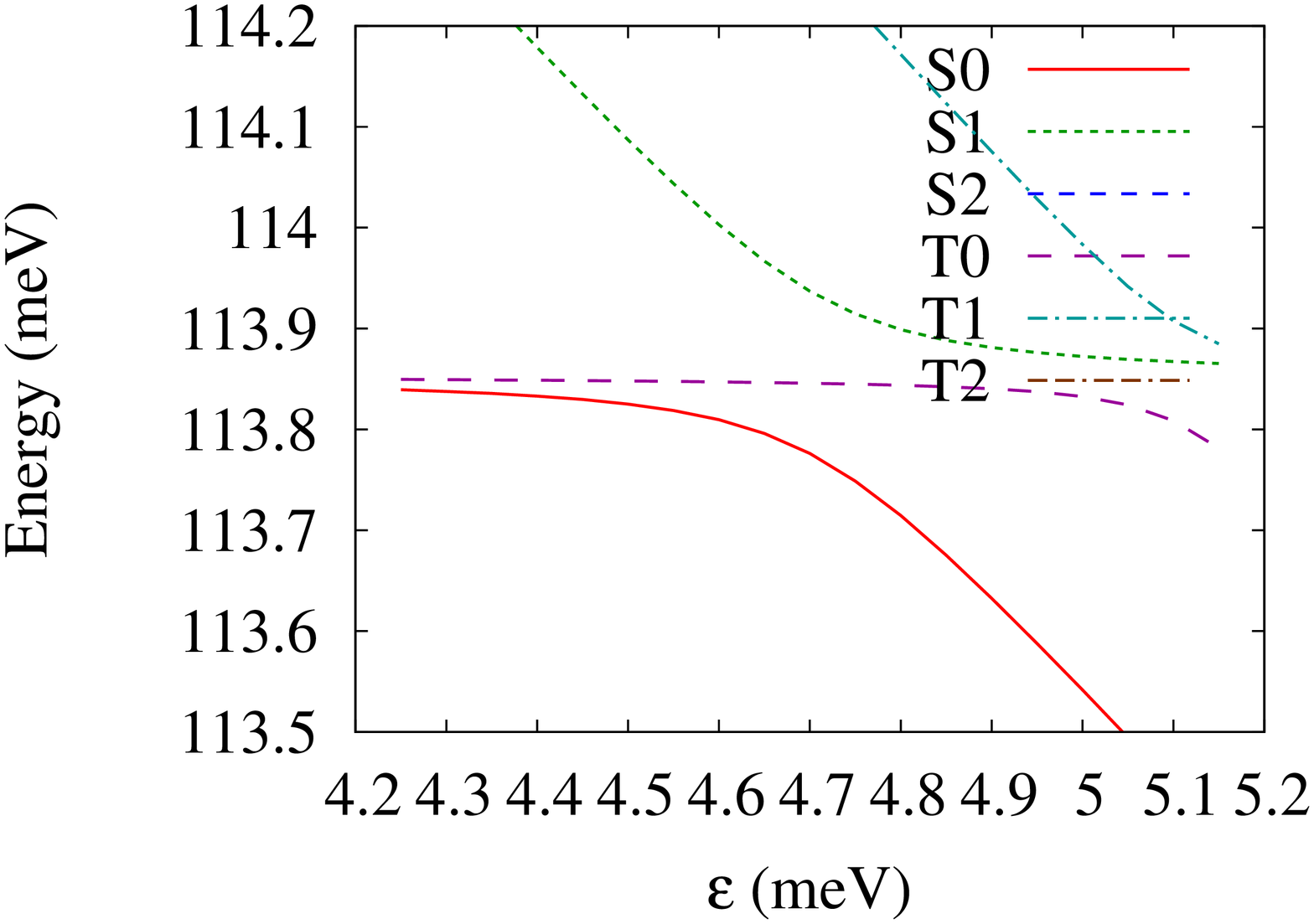}
\includegraphics[width=1in,angle=270]{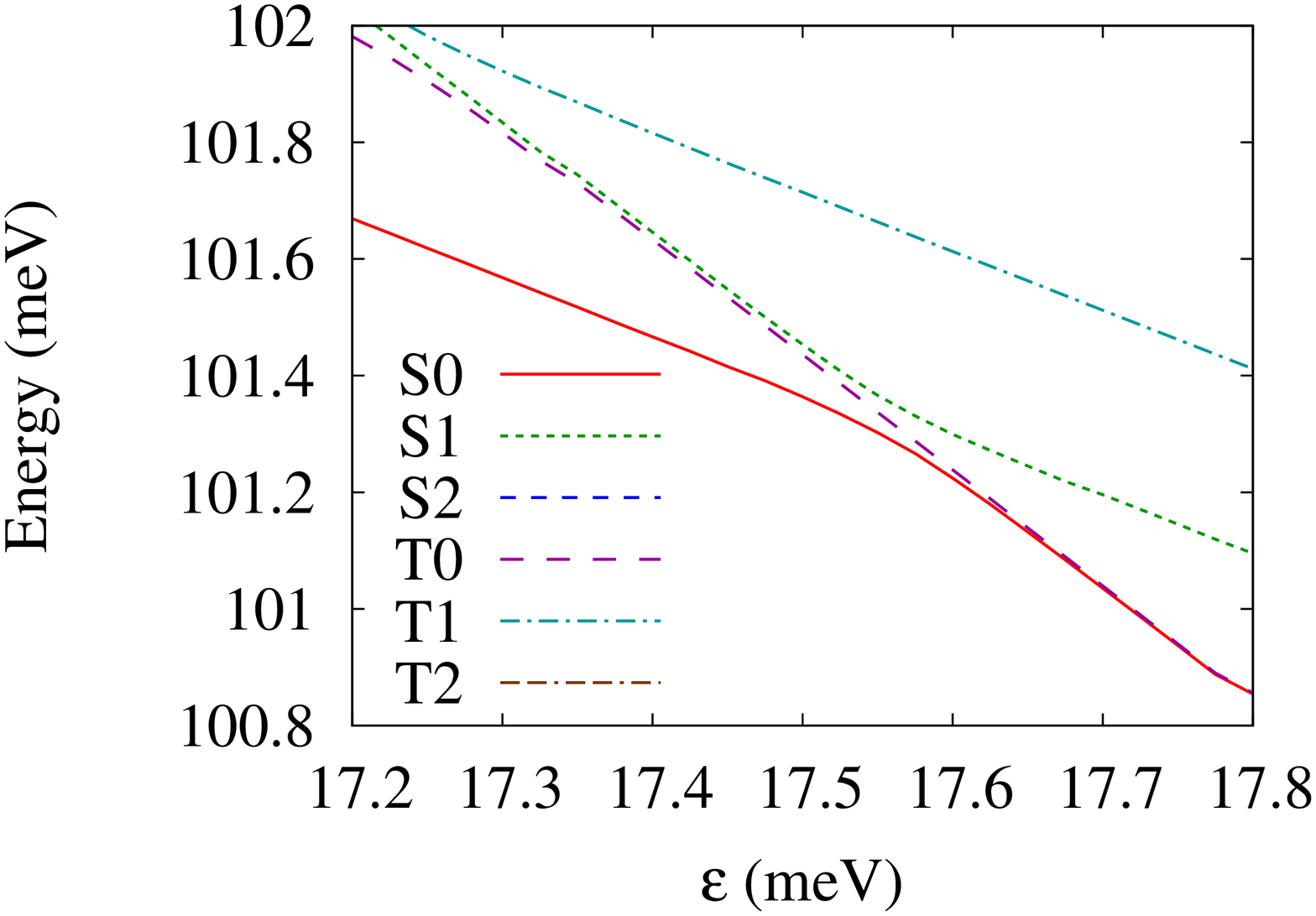}
\includegraphics[width=1in,angle=270]{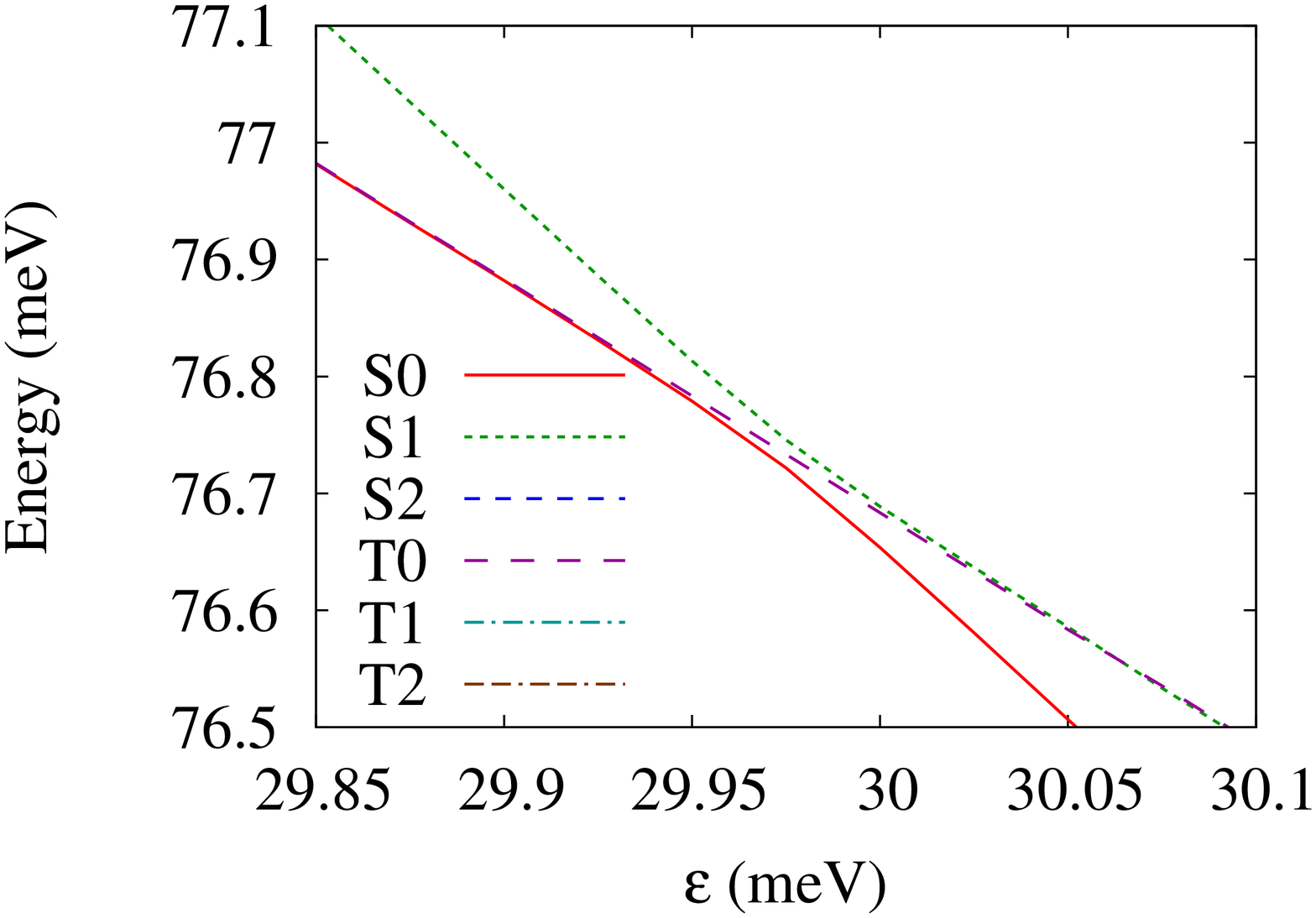}
\caption{6 electron energies vs detuning using CI-2 ($n_G=18$, $E_c=55\meV$) for medium-dot parameters, zoomed to each of the three anti-crossings, (3,3)-(2,4), (2,4)-(1,5), and (1,5)-(0,6) of the lowest singlet state.  $S_z=0$ for all states, and labels denote spin (S=singlet, T=triplet) and energy-ordering index.\label{fig6eSpectrumMediumZooms}}
\end{center}
\end{figure}

Previously we considered dots with the confinement energy $E_0=24.26\meV$, which corresponds to a ground state electron wave function (in GaAs) with diameter approximately $13\nm$.  In the medium-dot and large-dot parameter cases we now consider, the ground state wave function diameters are approximately $25\nm$ and $50\nm$ respectively.  The energy spectrum for the medium-dot DQD parameters is shown in Fig.~\ref{fig6eSpectrumMedium}, with zoomed-in views of the singlet anti-crossings in Fig.~\ref{fig6eSpectrumMediumZooms}.  The spectrum is very similar to the small-dot case.  There exists a low-lying isolated singlet-triplet subspace with avoided crossings, with an meV-scale gap to excited singlet and triplet states. The singlet-triplet gap between ground states in the (2,4) region is small and negative, but can be tuned to a large positive value by the magnetic field, as will be discussed in Section IV below.  The (0,6) exchange energy gap is positive and sufficiently large to enable proper qubit initialization.

\begin{figure}
\begin{center}
\includegraphics[width=2in,angle=270]{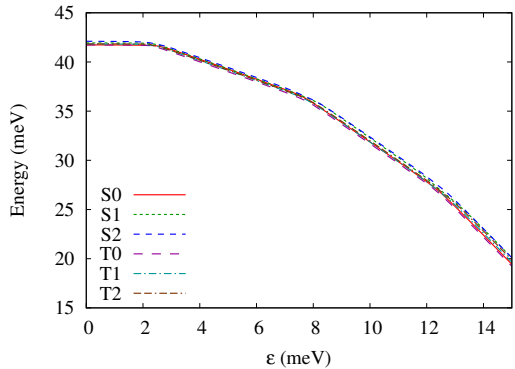}
\caption{6 electron energies vs detuning using CI-2 ($n_G=10$, no cutoff) for large-dot parameters ($E_0=1.82\meV$, $L=50\nm$, $B=212\mT$). $S_z=0$ for all states, and labels denote spin (S=singlet, T=triplet) and energy-ordering index. \label{fig6eSpectrumLarge}}
\end{center}
\end{figure}

\begin{figure}
\begin{center}
\includegraphics[width=1in,angle=270]{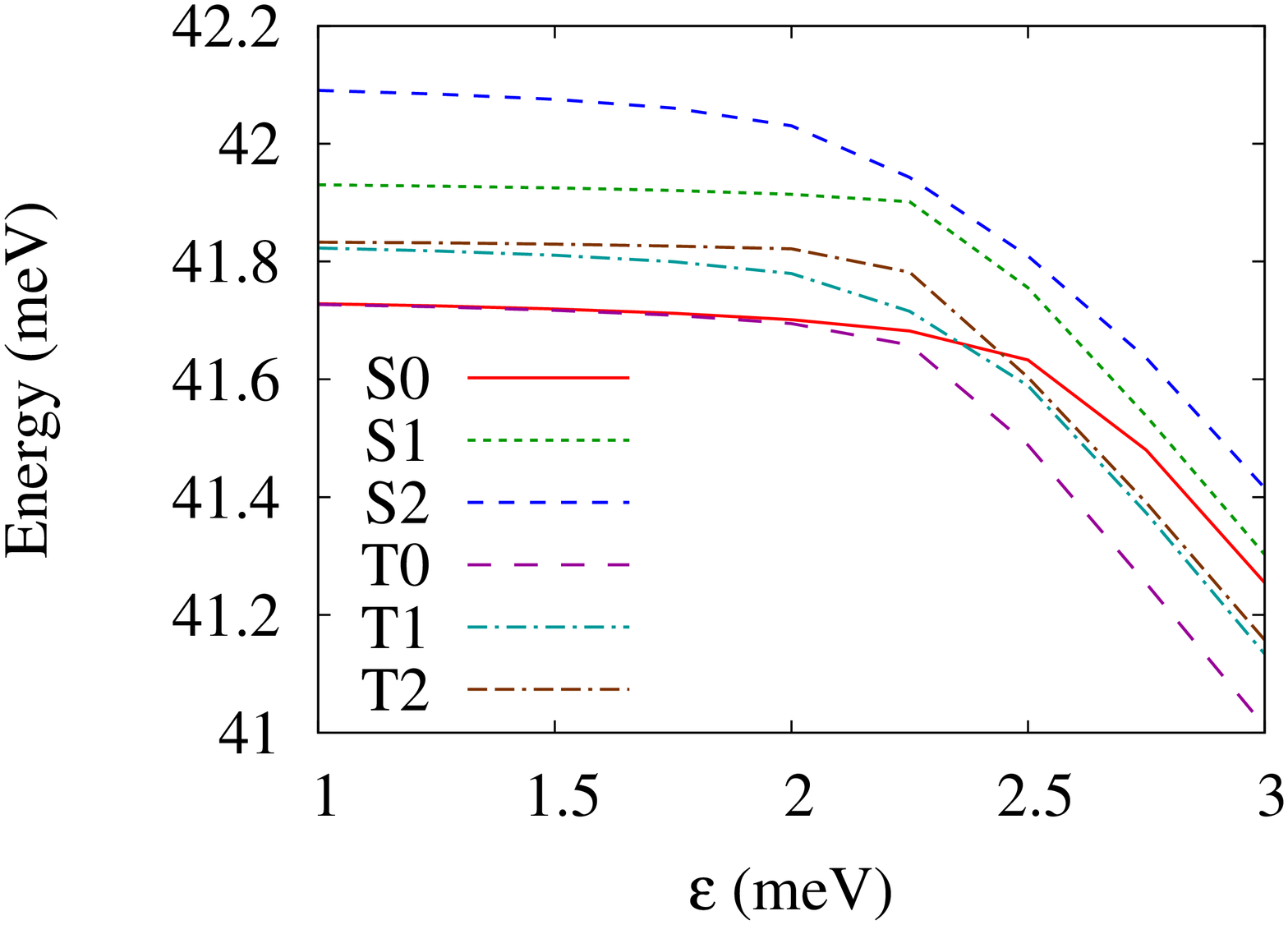}
\includegraphics[width=1in,angle=270]{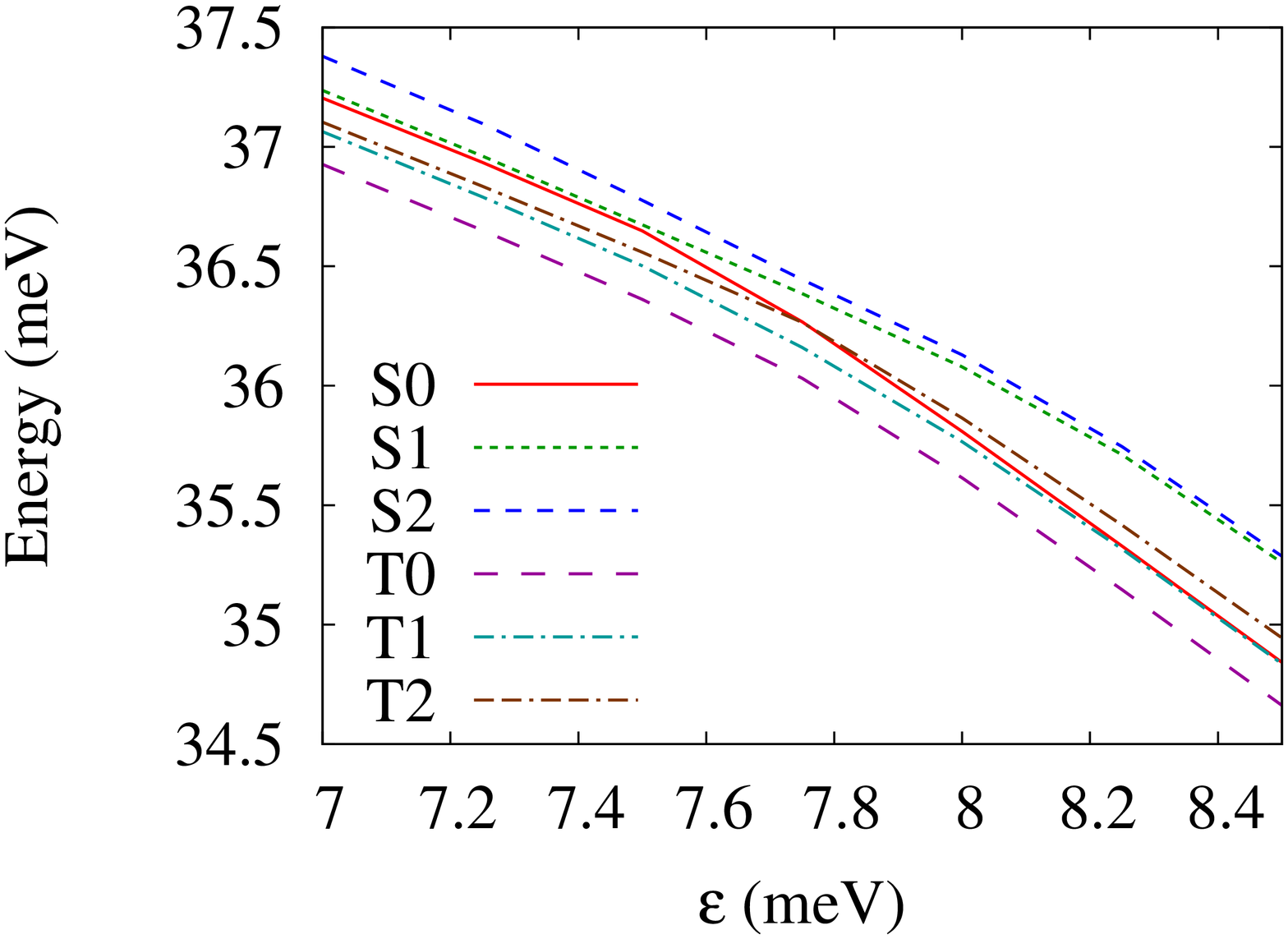}
\includegraphics[width=1in,angle=270]{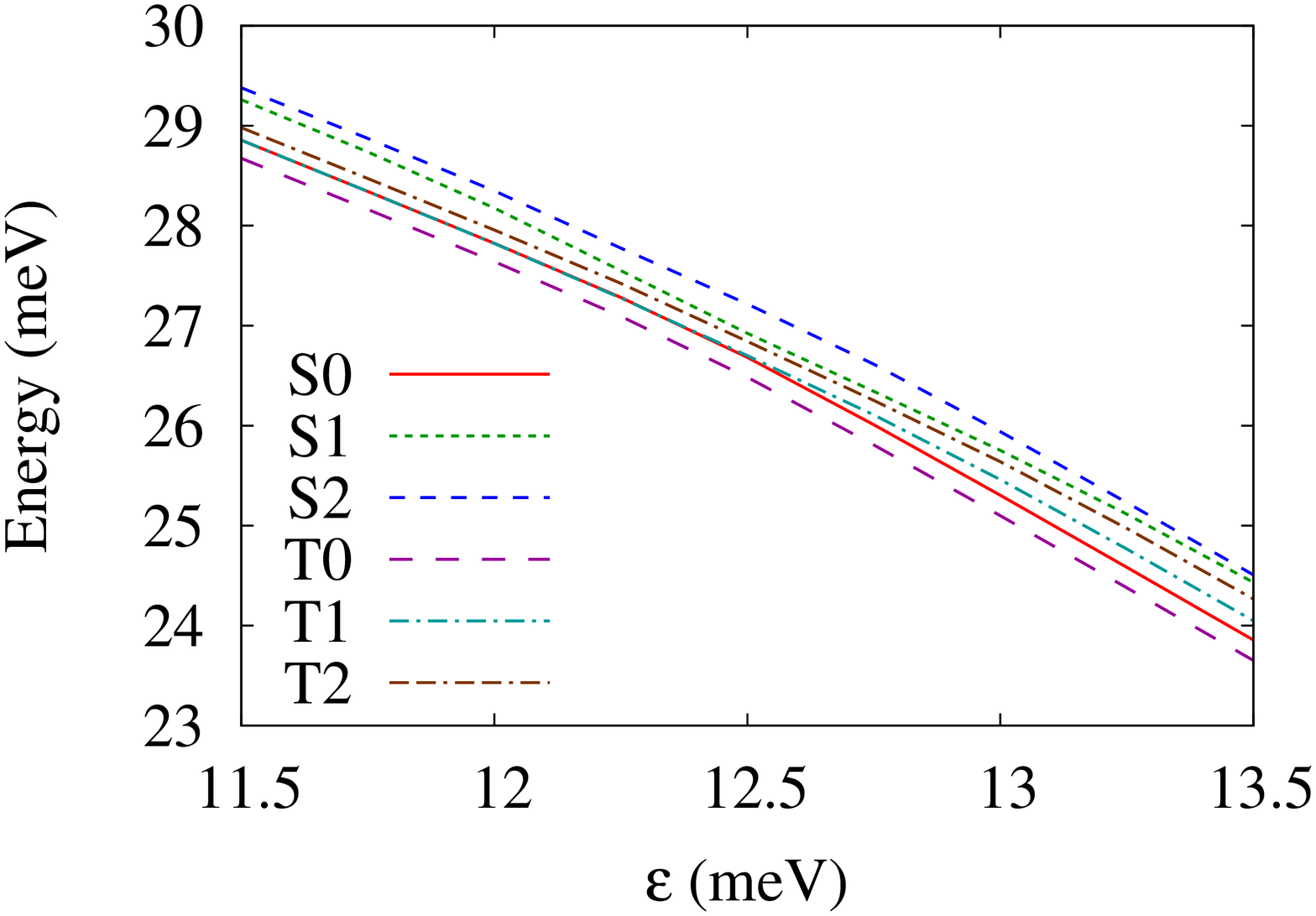}
\caption{6 electron energies vs detuning using CI-2 ($n_G=10$, no cutoff) for large-dot parameters, zoomed to each of the three anti-crossings, (3,3)-(2,4), (2,4)-(1,5), and (1,5)-(0,6) of the lowest singlet state. $S_z=0$ for all states, and labels denote spin (S=singlet, T=triplet) and energy-ordering index.\label{fig6eSpectrumLargeZooms}}
\end{center}
\end{figure}

Moving to even larger dots with $E_0=1.82\meV$ brings us into a regime where the CI technique is no longer fully convergent.  In the previous two cases where $E_0=24.26\meV$ and $7.28\meV$, the CI is able to converge with respect to its truncation, and in the $E_0=7.28\meV$ case we get qualitative convergence with respect to the Gaussian basis size (see Appendix \ref{ConvergenceAppendix}).  In particular, adding more Gaussian elements may lower the state energies a small amount but does not cause level reordering.   In the present case of large-dot DQDs ($E_0=1.82\meV$), this is not the case, and special care must be taken in the interpretation of the results. The numerical challenge stems from the fact that a large number of single-particle states are required to capture even the lowest energy multi-particle states, so that less truncation can be applied and we are therefore limited to smaller Gaussian basis sizes.  This limitation, in turn, restricts the variational freedom of the many-particle states, leading to less certainty in qualitative features such as the relative spacing and ordering of energy levels.  In the present case, computational power limits us to a Gaussian basis of size $n_G=10$ if we require complete convergence with respect to truncation.

\begin{figure}[h]
\begin{center}
\includegraphics[width=2.2in,angle=270]{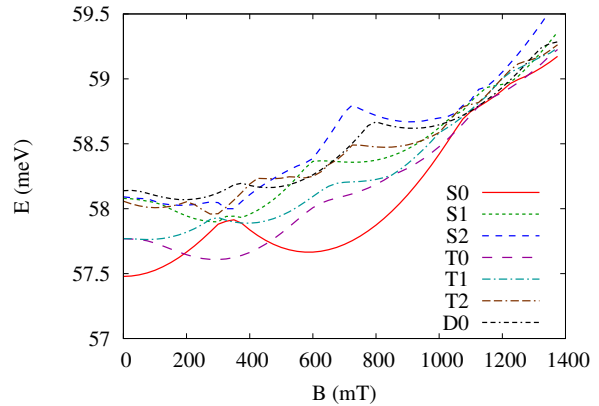}
\caption{Energy spectrum of a single dot with 6 electrons. Dot confinement energy $E_0=1.82\meV$.  At a vertical magnetic field $B=212\mT$, the spectrum approximates the (0,6) regime of a DQD with our large-dot parameters.  Computed using CI-2 with $n_G=13$ and no energy cutoff. $S_z=0$ for all states, and labels denote spin (S=singlet, T=triplet) and energy-ordering index.\label{figSingleDotJ_vs_B}}
\end{center}
\end{figure}

The resulting spectrum is shown in Fig.~\ref{fig6eSpectrumLarge}, with magnifications near the anticrossings shown in Fig.~\ref{fig6eSpectrumLargeZooms}. As expected, we see the continued trend of smaller energy spacings in the low-energy spectrum. We also note that a non-singlet state is the lowest-lying state in the (2,4), (1,5), and (0,6) charge sectors for $n_G=10$, and that such a situation presents a potential problem for initialization as discussed above. A non-singlet ground state is plausible since decreased orbital energy separations can make it favorable for electrons to occupy higher orbitals and form non-singlets in order to reduce Coulomb repulsion.  This line of reasoning is verified deep in the (0,6) region of the spectrum, where we can approximate the DQD system as a single dot with six electrons and solve just a single dot problem using the CI.  This allows us to place more Gaussians in the single dot and ultimately gives us a converged CI spectrum, which is shown in Fig.~\ref{figSingleDotJ_vs_B}.  We see that at $B=212\mT$ a triplet state is the ground state. However, it is important to note that if the magnetic field were reduced below $200\mT$ or increased above $500\mT$, the ground state will become a singlet, so that the ability to initialize the qubit can be restored by tuning the magnetic field appropriately.

Another observation that can be made regarding the DQD spectrum shown in Fig.~\ref{fig6eSpectrumLargeZooms} is that there \emph{may} be intervening levels between the two qubit states.  For instance, in the (2,4) region the lowest three states are all triplets, while the lowest singlet state is the third excited state. Thus if we were to use the lowest singlet and triplet states as the qubit space, there would be two intervening (triplet) levels in the (2,4) region. This would make it difficult to operate the qubit in the (2,4) region (a possibility the details and benefits of which will be elaborated on in Section IV), unless it is possible to tune the magnetic field such that the qubit states form a well separated ground space in (2,4) while maintaining the singlet as the ground state in (0,6). If this is not possible, then one could still restrict qubit operation to the (3,3) region, tuning through (2,4) solely for the purposes of initialization and readout.

\section{Qubit Characterization and Control\label{secCharacterization}}

\subsection{Similarities between 6- and  2-electron DQDs}
The control of a 6-electron singlet-triplet qubit is nearly identical to the 2-electron case.  By tilting the DQD, the exchange energy $J$ (the difference between the lowest triplet and singlet energies) can be tuned to cause qubit $z$-rotation, and an existing or applied gradient in the magnetic field causes an $x$-rotation.  Initialization is performed by biasing the qubit so that there is a large gap between the singlet and triplet states and waiting for the DQD to relax to the ground state.  Read-out is done by biasing the qubit so that the singlet and triplet are in different charge sectors and measuring the charge sector via a nearby charge sensor (usually a quantum point contact or a single-electron-transistor). 

\begin{figure}[h]
\begin{center}
\includegraphics[width=2in]{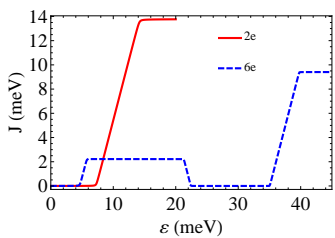}
\caption{Comparison of the exchange energy vs detuning in 6- and 2-electron DQDs using small-dot parameters (given in Table \ref{tblParameters}).  Note that the dot-separation parameter $L$ is different for the two cases in order to equate the valence tunneling of the 6- and 2-electron DQDs as described in Sec.~\ref{secResults}A.  We find comparable orders of magnitude, indicating that z-rotations of this 6-electron DQD qubit can be controlled in a similar manner to the rotation in the corresponding 2-electron DQD qubit. Calculations were performed with CI-1 using $E_c=30\hbar\omega_c/2=73.15\meV$ \label{figExchangeSmall}}
\end{center}
\end{figure}

The exchange energy as a function of tilt for our small-dot DQD parameters is shown in Fig.~\ref{figExchangeSmall}.  The qualitative shape of this curve is generic for 6e- DQDs with well separated dots:  there are two plateaus corresponding to the regions where both singlet and triplet are in the (2,4) or (0,6) charge sector, interspersed by two flat ``valleys'' in the (3,3) and (1,5) regions.  The heights of the (2,4) and (0,6) plateaus are related to the single-dot exchange energies for dots with 2, 4, and/or 6 electrons, and the sharpness of the ascent to and descent from the plateaus are a function of the inter-dot tunneling amplitude.  Larger tunneling (a lower barrier) results in greater spacing between the singlet and triplet avoided crossings at a given charge sector transition, and thus a more gradual ascent/descent.  The relatively small exchange energies of the (3,3) and (1,5) regions are determined by the inter-dot coupling. These features are analogous to the 2e case apart from the fact that there are two plateaus instead of one (as there are four charge sectors instead of two).  The possible advantages due to the existence of two plateaus will be discussed below.

\begin{figure}[h]
\begin{center}
\includegraphics[width=1.5in,angle=270]{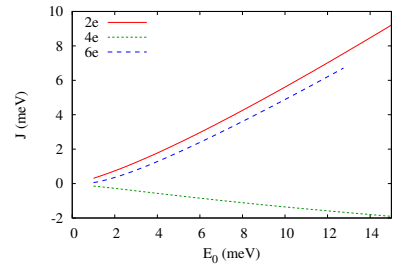}
\caption{Exchange energy of a single dot with confinement energy $E_0$ in magnetic field $B=847\mT$ (the same value used in our set of medium-dot DQD parameters, see Table \ref{tblParameters}) for 2, 4, and 6 electrons.  Computed using CI-2 with $n_G=13$ and no energy cutoff.\label{figSingleDotJ_vs_E0}}
\end{center}
\end{figure}

We also compare with a two-electron result in Fig.~\ref{figExchangeSmall}.  The two-electron DQD has a different DQD potential ($L=16.5\nm$ instead of $20\nm$), in order to match the valence tunneling of the six-electron DQD as described above.  We see that the magnitude of the six-electron DQD exchange energies is similar to that of the two-electron DQD.  Although it is computationally time-consuming to repeat the calculations across the whole range of dot sizes, it is easy to show that the similarity between six-electron and two-electron DQD exchange plateaus is independent of dot size by plotting the exchange energy of a \emph{single} dot as a function of its confinement energy.  When the inter-dot coupling is small, the heights of the double-dot exchange plateaus are essentially the exchange energies of a single-dot at the appropriate filling, so that the 2-electron (0,2) and 6-electron (0,6) plateau heights correspond to the two- and six-electron exchange energies of a single dot. Figure \ref{figSingleDotJ_vs_E0} thus implies the similarity of 6-electron (0,6) and 2-electron (0,2) DQD exchange plateaus over a large range of confinement energies. We can also estimate the dependence of the (2,4) plateau height based on dot confinement energy when (1) the 2- and 4-electron single dots each have an isolated singlet-triplet low energy subspace, and have substantially different exchange energies,  and (2) the dots are weakly coupled.  In this particular case, the lowest 6-electron singlet or triplet states can be approximated by a singlet in the dot with the larger exchange energy and a singlet or triplet, respectively, in the dot with the smaller exchange energy.  For $B=847\mT$, the case of Fig.~\ref{figSingleDotJ_vs_E0}, the 2-electron exchange energy is only somewhat larger than the 4-electron exchange energy (which is negative due to the degeneracy of the first excited dot orbital -- Hund's rule), and we see that while the (2,4) 6-electron exchange energy is also negative, it's magnitude is much smaller than the 4-electron single-dot exchange energy, indicating the 6-electron singlet and triplet have excited state components beyond the simple picture outlined above.

\begin{figure}[h]
\begin{center}
\includegraphics[width=2.2in]{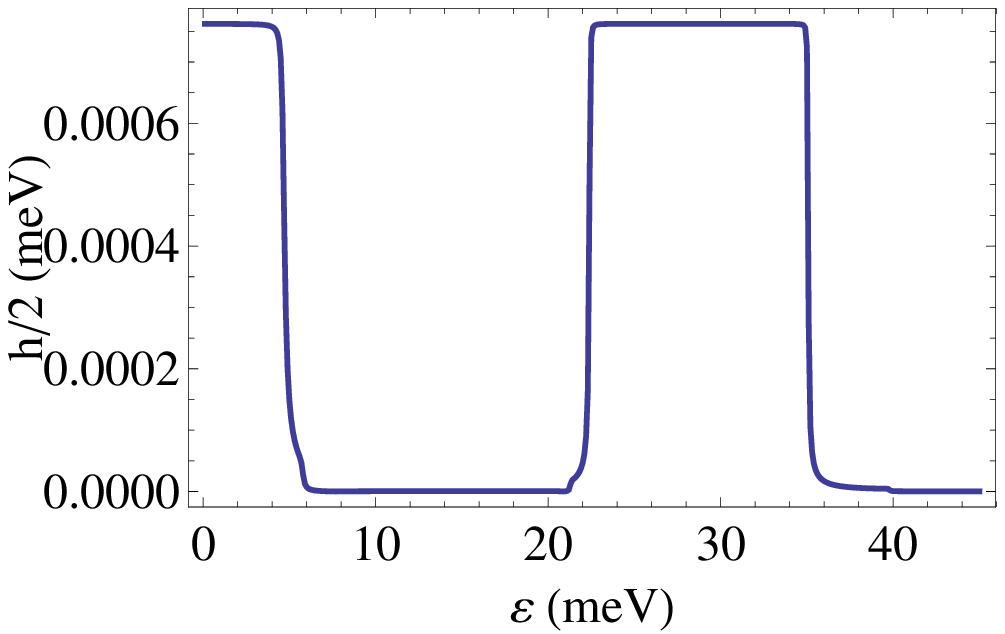}
\includegraphics[width=2.2in]{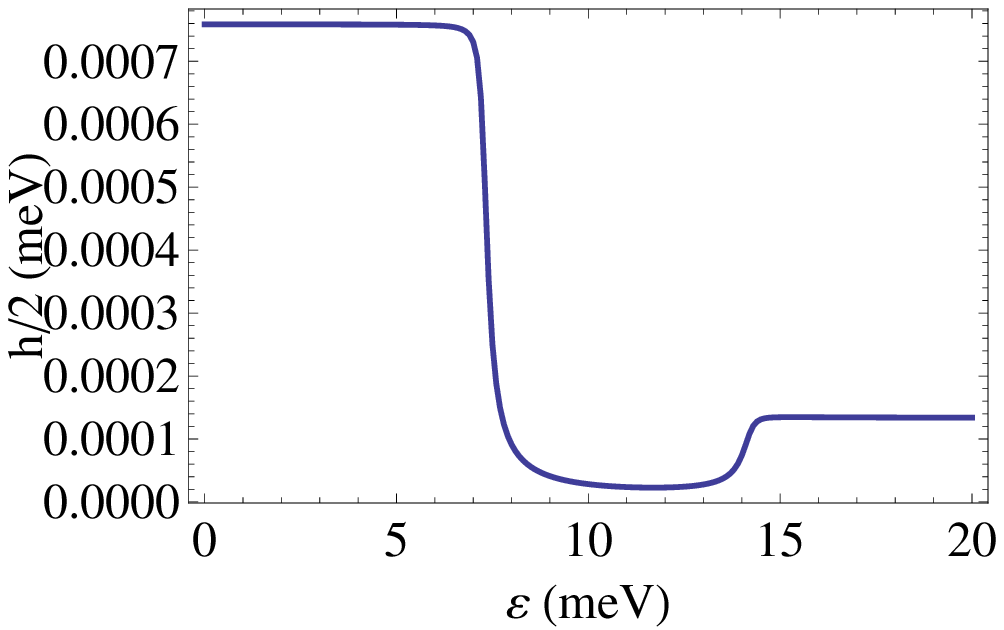}
\caption{Comparison of the singlet-triplet energy coupling, $h$, in 6-electron (top pane) and 2-electron (bottom pane) DQDs using small-dot parameters (given in Table \ref{tblParameters}). We find comparable order of magnitudes, indicating that x-rotations of this 6-electron DQD qubit can be controlled in a similar manner to the rotation in the corresponding 2-electron DQD qubit.Calculations were performed with CI-1 using $E_c=30\hbar\omega_c/2=73.15\meV$.\label{figHz12Small}}
\end{center}
\end{figure}

\begin{figure}[h]
\begin{center}
\includegraphics[width=2.2in]{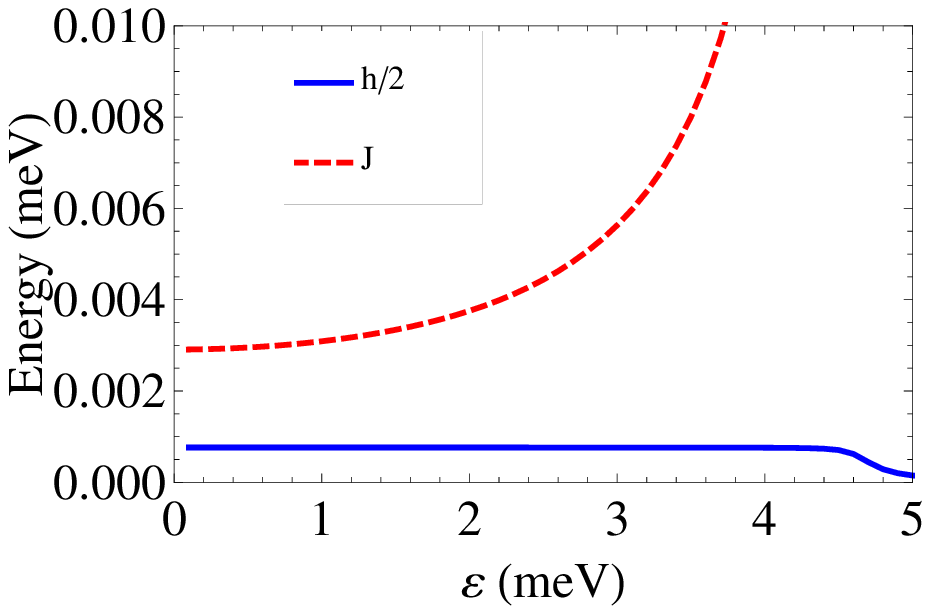}
\includegraphics[width=2.2in]{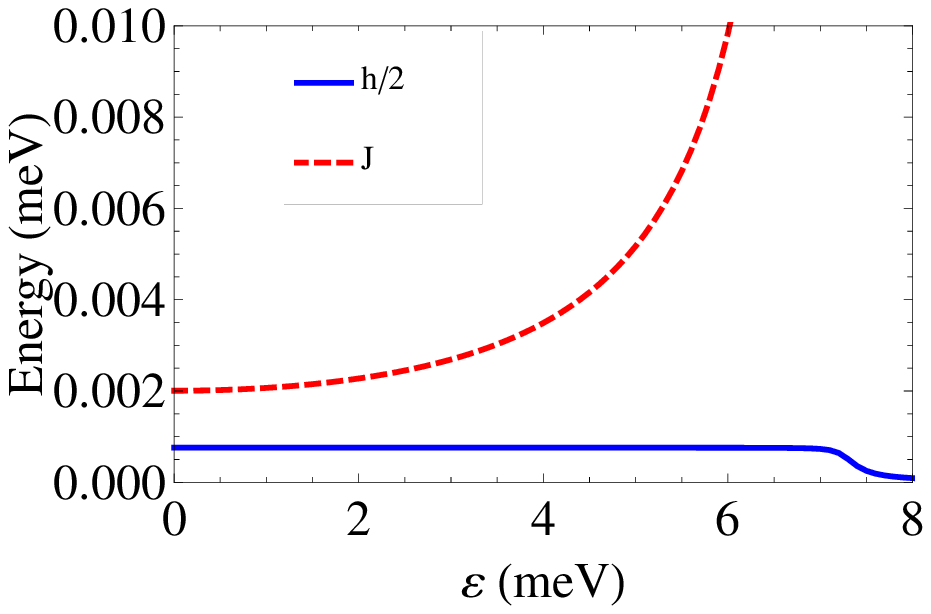}
\caption{Comparison of the singlet-triplet energy coupling $h$, and exchange energy $J$, vs detuning in the (3,3) region of a 6-electron DQD (top pane) and the (1,1) region of the corresponding 2-electron DQD (bottom pane) using small-dot parameters (given in Table \ref{tblParameters}).  The similarity indicates that this 6-electron DQD could be controlled as a qubit in the same way as the corresponding 2-electron DQD.
\label{figHz12JSmall}}
\end{center}
\end{figure}

In order to design qubit operations, it is very helpful to construct an effective Hamiltonian describing the two-level ground space. In the case of our multi-electron DQDs, this Hamiltonian assumes the following form:
\begin{equation}
H_{eff}=\frac{J}{2}\sigma_z+\frac{h}{2}\sigma_x,   
\end{equation}
where $\sigma_x$, $\sigma_z$ are Pauli matrices, $J$ is the exchange splitting between the lowest two levels $|g\rangle$ and $|e\rangle$, and a magnetic field gradient with Zeeman energy $h/2=\langle g|\frac{g^*\mu_B\Delta B}{2L}x|e\rangle$ couples these states, with $\Delta B=B_R-B_L$ denoting the difference between the magnetic fields in the right and left dots. In the case of two-electron singlet-triplet qubits, field gradients on the order of tens to hundreds of milliTesla have been generated and stabilized via a feedback mechanism involving repeated driving
through singlet-polarized triplet anticrossings \cite{Foletti_NP09,Bluhm_PRL10}. Using CI-1, we have computed the Zeeman energy as a function of detuning for both the two-electron and six-electron cases with $\Delta B=60$ mT, as
shown in Fig.~\ref{figHz12Small}. It is apparent from Fig.~\ref{figHz12JSmall} that the six-electron values of $J$ and $h$ in the (3,3) region are comparable to those in the (1,1) region of the two-electron case (both are on the order of $\mu$eV). Therefore, the effective two-level Hamiltonians for two and six electrons are qualitatively the same, so that qubit rotations can be implemented using similar procedures in the two cases. For example, recent
progress in developing robust singlet-triplet qubit rotations in two-electron DQDs \cite{Wang_NatComm12,Kestner_arxiv13} should apply equally well to the six-electron case.

\begin{figure}[h]
\begin{center}
\includegraphics[width=2.2in]{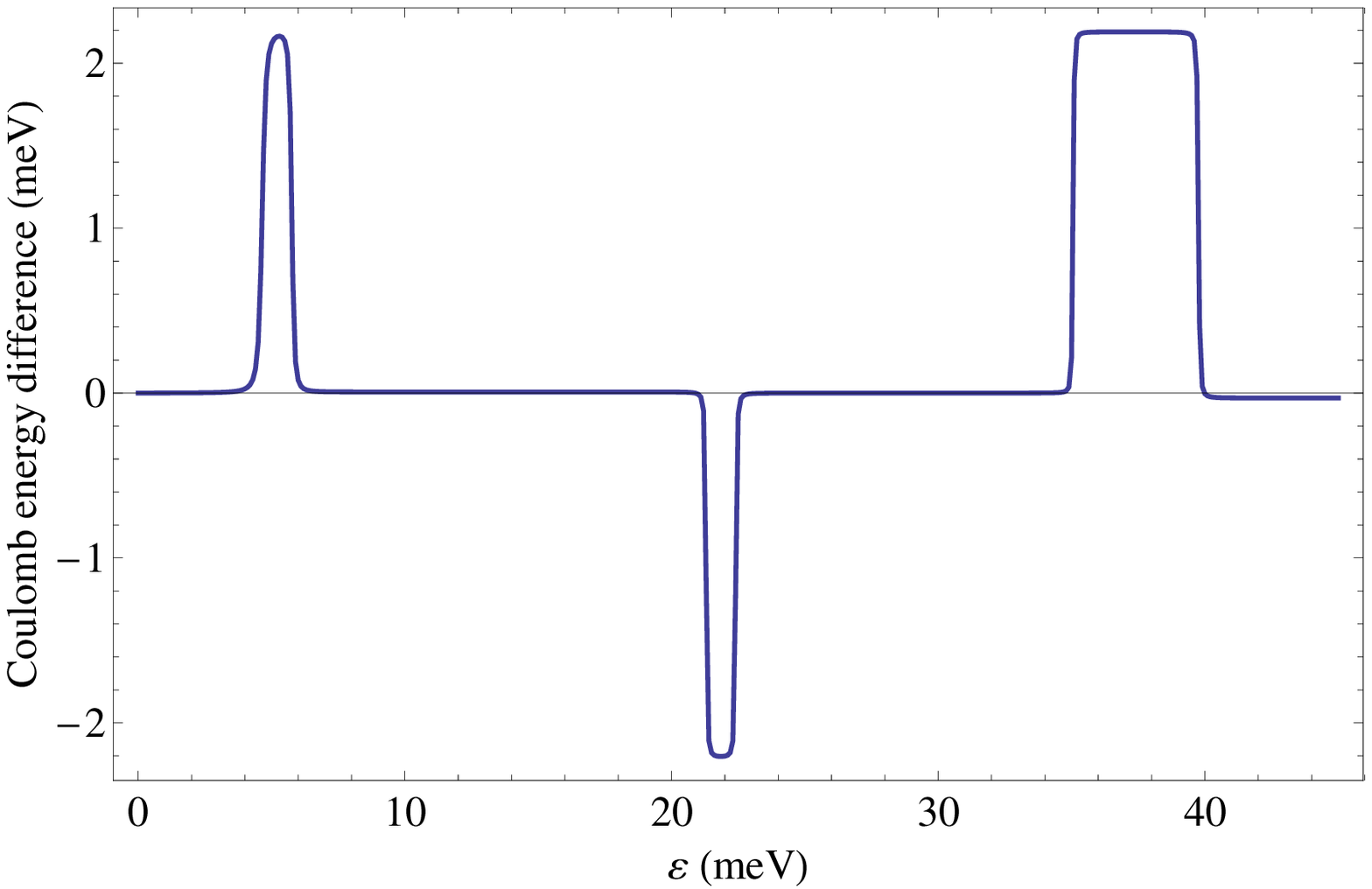} 
\includegraphics[width=2.2in]{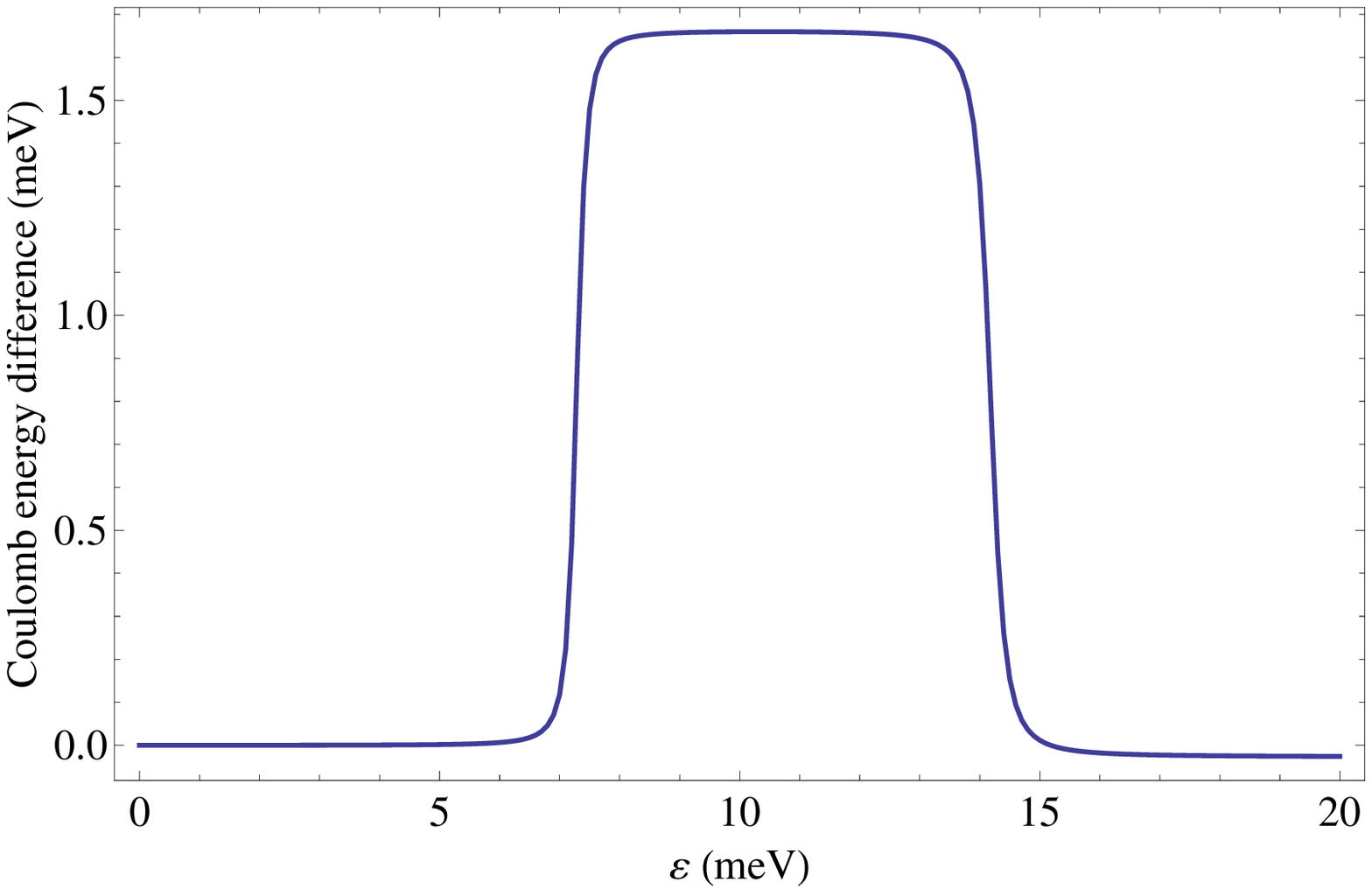} 
\caption{Comparison of the difference in Coulomb energy (defined as $\int e^2|\Psi(r)|^2/(r-r_0)\,dr$) between the singlet and triplet states 50nm away from the DQD for the 6-electron (top) and 2-electron (bottom). Small-dot parameters are used as given in Table \ref{tblParameters} and computation was done with CI-1 using $E_c=30\hbar\omega_c/2=73.15\meV$.\label{figCoulombEnergySmall}}
\end{center}
\end{figure}

Qubit read-out is performed via a nearby charge sensor, and relies on the sensitivity of the charge sensor to the DQD state.  Many factors enter into the coupling between the DQD charge state and the signal from a charge sensor, and we do not account for many of these here.  Instead we consider the difference in Coulomb energy of the singlet and triplet states at a point in space $50\nm$ away from a DQD (this distance is arbitrary, and is meant as an estimate of the location of a charge sensing constriction or dot).  We define Coulomb energy to be the integral $\int e^2|\Psi(r)|^2/(r-r_0)\,dr$, where $\Psi$ is the wave function whose Coulomb energy we're computing (the lowest singlet or triplet state) and $r_0$ is the point in space $50\nm$ away from the DQD.  Thus, the difference in the Coulomb energy of the singlet and triplet states at the location of a charge sensor gives an indication of how strongly the DQD state couples to the sensor (or equivalently, how easy it is to perform read-out on the qubit).  We find that the ease of read-out is similar for 2-electron and 6-electron DQDs (using our small-dot parameters), shown in Fig.~\ref{figCoulombEnergySmall}.

Clearly, in order to use the (2,4) or (0,6) regions for readout or initialization requires adiabatically sweeping through avoided crossings in the spectrum (see, e.g., Fig.~\ref{fig6eSpectrumMediumZooms}).  By the Landau-Zener formula for an avoided crossing with minimum energy gap $2\delta$, the rate at which the energy gap is changed, $v$, should be less than $\delta^2/\hbar$ to avoid the diabatic transition that occurs with probability $e^{-2\pi \delta^2/\hbar v}$.  Thus, an adiabatic sweep across an anticrossing starting from a detuning with a larger energy gap $2\Delta$ requires ramping time $T > \hbar\Delta/\delta^2$.  Even for $\Delta \sim 2\delta \sim 100\mu$eV, ramp times of tens of ps satisfy this condition.  On the other hand, if the ramp is too slow the magnetic field gradient will rotate the singlet toward the triplet during the sweep.  For typical field gradients less than a $\mu$eV, this prohibits sweeps longer than about a ns.  So, sweeping the detuning for readout and initialization should not pose a problem for ramping times on the order of 100 ps.  In the event it does become difficult to satisfy the above conditions, nonlinear ramps provide an obvious alternative approach.

\subsection{Differences between 6- and 2-electron DQDs}
Along with the many similarities just discussed, there are some important differences between 6-electron and 2-electron singlet-triplet DQD qubits.  One, of course, is the difference that motivates this work: more electrons result in an increased ability to screen charge impurities, as investigated in Ref.~\onlinecite{Barnes_Kestner_Screening_2011}.  The incremental increase in screening ability with the number of filled shells is found to fall off at high electron number, but there is clear benefit from the increase to 6-electrons (1 filled shell) from 2-electrons (no filled shells) in a DQD.  This may mean that 6-electron DQD qubits are less susceptible to charge noise and the random potential caused by disorder, making them easier to realize from both a materials and charge-noise standpoint.  Additionally, if the number of electrons could be inferred without emptying the DQD completely and counting up charge transitions (as it is typically done at present), a 6-electron DQD could prove easier to fabricate from a lithography standpoint.


On the other hand, although the energy scales are acceptable in the 6-electron DQDs, the low-lying qubit subspace is less isolated from the higher levels than in the 2-electron case, particularly when dealing with large dots.  This is the only disadvantage we see to using 6-electron DQD qubits, and we believe it is far outweighed by the benefits afforded.  Clearly, though, one cannot increase \textit{ad infinitum} the number of filled shells to achieve increased robustness to charge impurities.  The many-body energy gaps are dependent on the dot size (confinement energy) and magnetic field in addition to the electron number, and what energy gaps can be tolerated will depend on temperature, qubit gate speed, and readout technique.  In short, precisely determining the optimal (odd) number of electrons per dot will be a non-universal system-dependent quantity.  With that said, we speculate that for typical GaAs and Silicon (assuming a large valley splitting) DQDs this optimal number will be somewhere between three and nine electrons per dot, since three appears to be clearly better than one, and by seven the charge-screening benefits are already seen to be minor (c.f. fourteen-electron DQD results using the frozen-core approximation in Ref.~\onlinecite{Barnes_Kestner_Screening_2011}). 

\begin{figure}[h]
\begin{center}
\includegraphics[width=2.2in,angle=270]{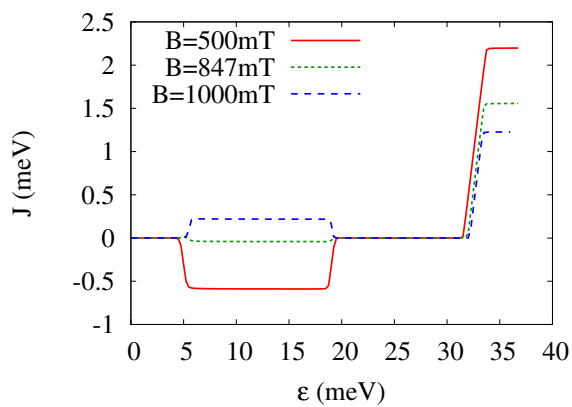}
\caption{Exchange energy vs detuning using CI-2 ($n_G=10$) for medium-dot parameters $E_0=7.28\meV$ and $L=50\nm$, and magnetic fields $B=500$, $847$, and $1000\mT$.  The ability to tune the \emph{two} exchange ``plateaus'' in the (2,4) and (0,6) regions with magnetic field is one possible advantage of a 6-electron DQD qubit over its 2-electron counterpart.\label{figExchangeMedium_varyB}}
\end{center}
\end{figure}

Another difference that is potentially very useful is that a 6-electron qubit has more charge sectors and hence more control capabilities than a 2-electron qubit.  There is a practical question of how much tuning range can be realized in an actual 6-electron device, but for the present discussion let us assume that tuning between all four charge sectors is possible.  Then there are two plateaus where $J$ is relatively small (typical parameters give $J$ ranging from nano- to micro-electron-volts), and two plateaus with potentially much larger $J$ (milli-electron-volts with typical parameters).  As we discussed above, for DQDs with well separated dots, the heights of these plateaus are of the same order as the exchange energies of single dots with appropriate numbers of electrons.  
The plateau heights depend on the confinement energy, as shown in Fig.~\ref{figSingleDotJ_vs_E0}, and the magnetic field. The $B$-field dependencies of the (2,4) and (0,6) plateaus are not proportional to each other, so the ratio of the plateau values can be tuned by $B$-field, as shown in Fig.~\ref{figExchangeMedium_varyB} at medium-dot DQD parameters.
The existence of regions where the exchange energy is largely insensitive to the detuning of the DQD can be used to rotate (or not rotate) the qubit in a way that is robust to detuning noise.  This is the long-standing idea behind the ``sweet spot" or ``park'' state, and has been studied\cite{Xuedong_ChargeDephasing_2006,StopaImmunity_2008,CulcerChargeNoise_2009,Nielsen_lowNoiseExchange_2010} for the case of rotation in a 2-electron DQD.  It was found that due to current limitations of controlling electronics (minimum voltage pulse widths), the value of the exchange flat must fall in the range of $1-10\mueV$, which is much smaller than the $O(\meV)$ exchange energy required for good initialization and read-out.\cite{Nielsen_lowNoiseExchange_2010}  In the 6-electron case, one can tune the system so that the one plateau is low ($1-10\mueV$) and one is high $O(\meV)$ in magnitude so that both operations can be done at a single magnetic field (see, e.g., Fig.~\ref{figExchangeMedium_varyB}).  Another new capability with 6 electrons is to use both (2,4) and (0,6) plateaus for rotations at different rates, or even different signs (again, see Fig.~\ref{figExchangeMedium_varyB}).  This could be leveraged to perform error correction to the extent that the environments seen by the (2,4) and (0,6) states are similar\cite{Khodjasteh09}.  


\section{Conclusions\label{secConclusions}}
We have shown that filling a DQD with 6-electrons results in a qubit that can be just as easy to initialize, manipulate, and read out as the well-established 2-electron qubit.  Large exchange energies in the (2,4) and (0,6) charge sector regions allow for initialization, and the exchange energy and singlet-triplet coupling via $B$-field gradient ($h$) dependencies are similar in the two cases, indicating qubit manipulation proceeds in the same way.  We find similar read-out sensitivity, since in both cases within the ``readout regions'' located between charge sectors there is a single electron charge transfer that differentiates the singlet and triplet states.  We find exchange energies and gaps to excited non-qubit-space states are of order tenths of milli-electron Volts ( $\gg k\mathrm{T}$ in a dilution refrigerator) in DQDs sized to match experimental capabilities.   Furthermore, 6-electron DQDs offer an enhanced charge screening  as well as richer control capability.  This work implies that the devotion of experimental resources to the realization of single-electron dots is unnecessary for, and even detrimental to, qubit performance.  In summary, comparison of 6-electron and 2-electron DQDs indicates that multi-electron quantum dots may be superior than single-electron dots for quantum computing purposes due to their additional controlability and intrinsic screening of unwanted charge impurities.  

Research was supported by LPS and by the Office of the Director of National Intelligence, Intelligence Advanced Research Projects Activity (IARPA), through the Army Research Office grant W911NF-12-1-0354, and by the Laboratory Directed Research and Development program at Sandia National Laboratories. Sandia National Laboratories is a multi-program laboratory managed and operated by Sandia Corporation, a wholly owned subsidiary of Lockheed Martin Corporation, for the U.S. Department of Energy's National Nuclear Security Administration under contract DE-AC04-94AL85000.

\appendix

\section{Convergence of CI-1 and CI-2\label{ConvergenceAppendix}}

\begin{figure}[h]
\begin{center}
\includegraphics[width=1.5in,angle=270]{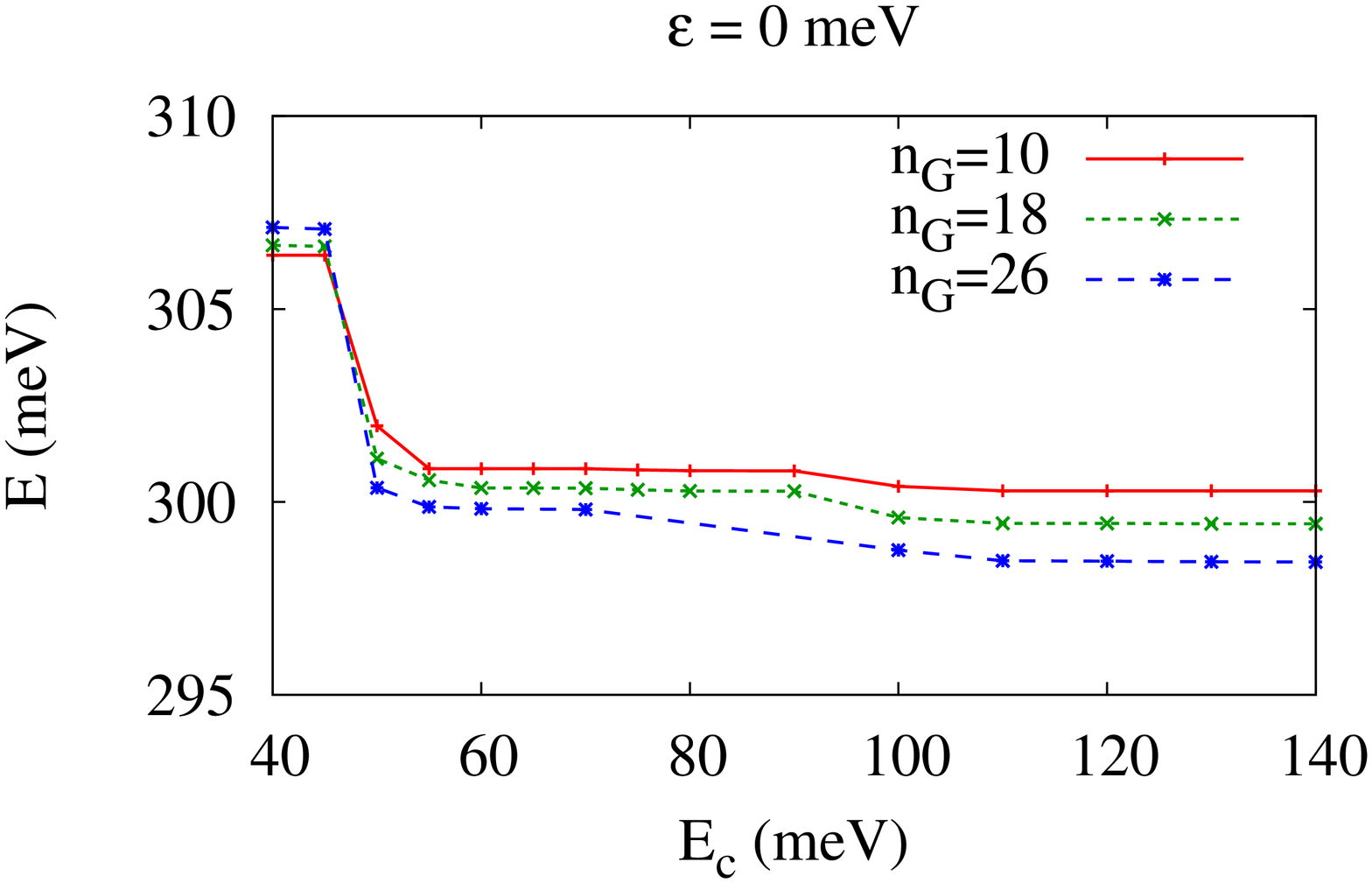}
\includegraphics[width=1.5in,angle=270]{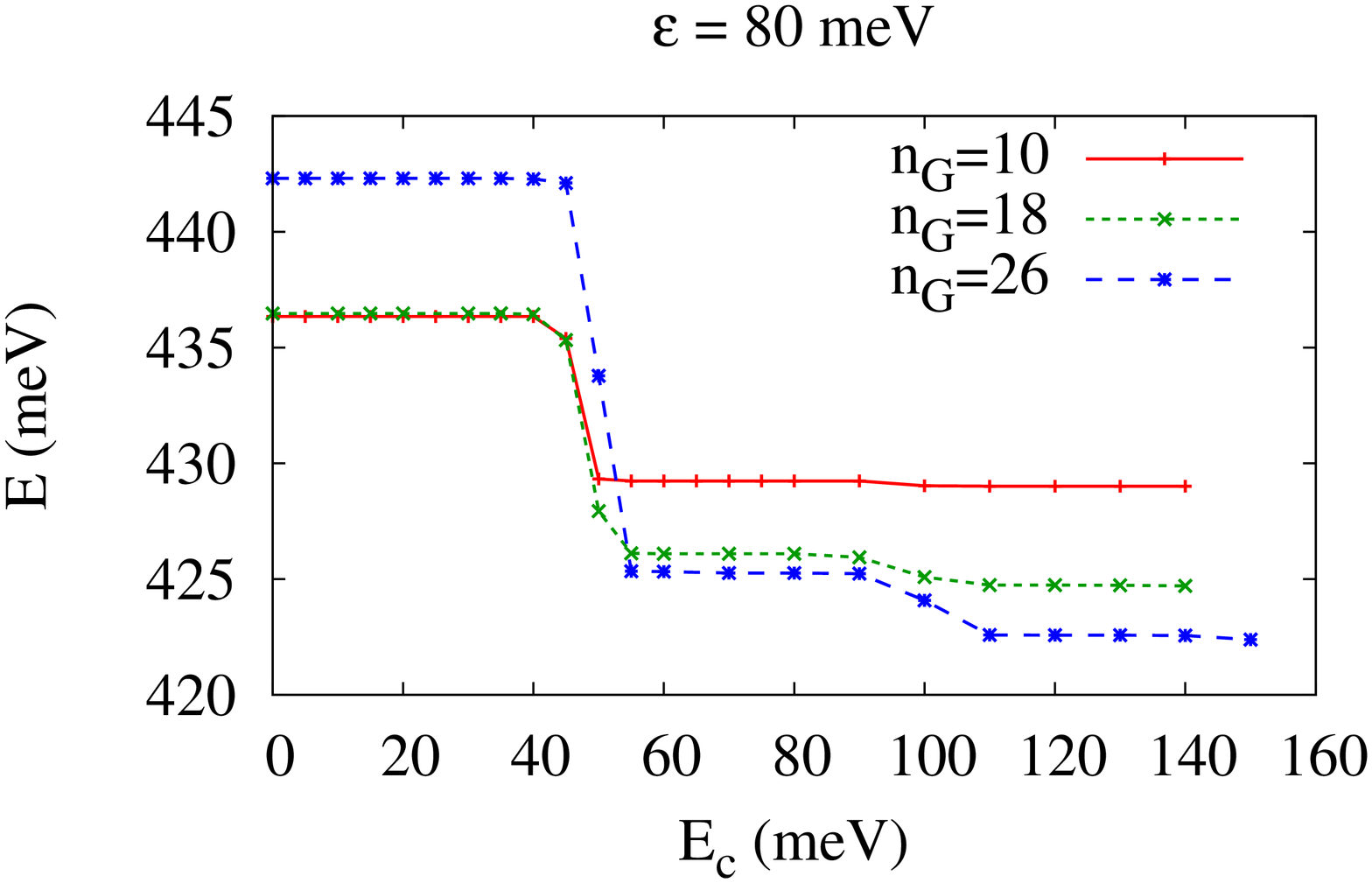}
\caption{Convergence of the lowest singlet energy (S0) of the CI-2 method for a 6-electron DQD in the (1,1) regime (upper pane) and (0,6) regime (lower pane).  Small dot parameters $E_0=24.26\meV$, $L=20\nm$, and $B=2.8\Tesla$, are used and the tilt $\epsilon$ is indicated above each plot.\label{figCI2convSmall}}
\end{center}
\end{figure}

We consider in this appendix the convergence of the CI-1 method as a function of the cutoff energy, and of the CI-2 method as a function of the number of Gaussians used and the cutoff energy. First, we consider the convergence behavior of 6-electron DQDs when using the small-dot parameters in Table \ref{tblParameters}.  The convergence of CI-1 with cutoff was shown above (Fig.~\ref{figConvergenceSmall}), and consists of steps at the energies where higher-shell Fock-Darwin orbital are introduced into the basis.  At $E_c=30\hbar\omega_c/2$, the value used to produce Figs.~\ref{fig6eSpectrumSmall} and \ref{fig6eSpectrumSmallZooms}, the single particle basis included the lowest ten states of each dot, consisting of Fock-Darwin S, P, D, and F states, and the many-body basis consisted of 1636 states.  In the case of CI-2, Fig.~\ref{figCI2convSmall} shows the convergence of the ground state energy within the (3,3) and (0,6) regions of detuning space.  The flattening of the curves at high $E_c$ indicates convergence with respect to the cutoff used.  Note, however, that even after results at a given basis size (fixed $n_G$) are converged with respect to the cutoff, increasing the number of Gaussian basis elements $n_G$ can continue to lower the energy of the ground state.  The disparity between the $E_c$-converged values at different $n_G$ is greater in the (0,6) detuning regime than in the (3,3).  This is a confirmation of what one may intuitively expect: that when all six electrons are in a single dot the additional variational degrees of freedom given by increasing the number of Gaussians per dot has more energetic impact.  In the (2,4) and (1,5) regimes we see convergence behavior intermediate between the (3,3) and (0,6) regions, and thus focus our convergence analysis to the (3,3) and (0,6) regimes.

\begin{figure}[h]
\begin{center}
\includegraphics[width=1.5in,angle=270]{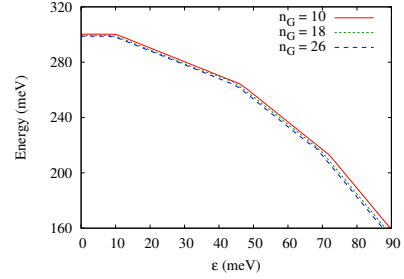}
\caption{Lowest singlet level vs detuning of a 6-electron DQD using CI-2 with different numbers of Gaussian basis elements $n_G$.  Small dot parameters are used: $E_0=24.26\meV$, $L=20\nm$, $B=2.8\Tesla$, and for each value of $n_G$ the energy is converged with respect to the cutoff parameter $E_c$. In the 10-Gaussian case, the Gaussian's centers are arranged 5 per dot in a "plus sign" pattern, with one Gaussian in the center of the dot and one along each of the +/-x and +/-y axis directions.  In the 18 Gaussian case, the centers are arranged in 3 by 3 grids on each dot, and in the 26-Gaussian case in addition to a 3 by 3 grid there is a Gaussian even further out along each of the +/-x and +/-y directions.\label{figCI2gaussianConv}}
\end{center}
\end{figure}

It initially appears concerning that the ground state energy is not converged with respect to the size of the CI's single particle basis, that is, the number of Gaussians.  While ideally one would like to see convergence in $n_G$ as well as $E_c$, we fine that this impossible from a computational standpoint and that the lack of convergence does not affect the qualitative aspects of the results.  Figure \ref{figCI2gaussianConv} provides an example of this, showing the 6-electron spectrum for $n_G=10$, $18$, and $26$ each converged with respect to $E_c$.  We see that all three curves show the same qualitative behavior over the entire range of detuning, with the larger $n_G$ curves lying slightly lower in energy.  Thus, while we require results to be converged with respect to the CI's cutoff energy we do not (and cannot, given practical computational limitations) require that the results are numerically converged with respect to the single particle basis size.  Instead, we content ourselves in a qualitative convergence with respect to the basis size and the layout of the Gaussians.  We remark that in the CI-1 method, the basis size is also fixed to a pre-defined number of Fock-Darwin orbitals, and that we make the same decision to prioritize convergence with respect to $E_c$ in that case, even though this is not as explicit in the text.  

\begin{figure}[h]
\begin{center}
\includegraphics[width=2in]{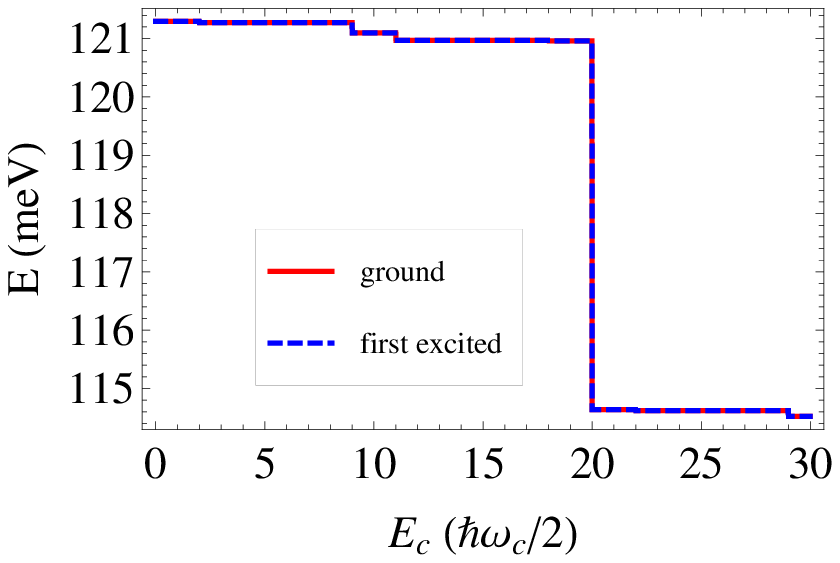}
\includegraphics[width=2in]{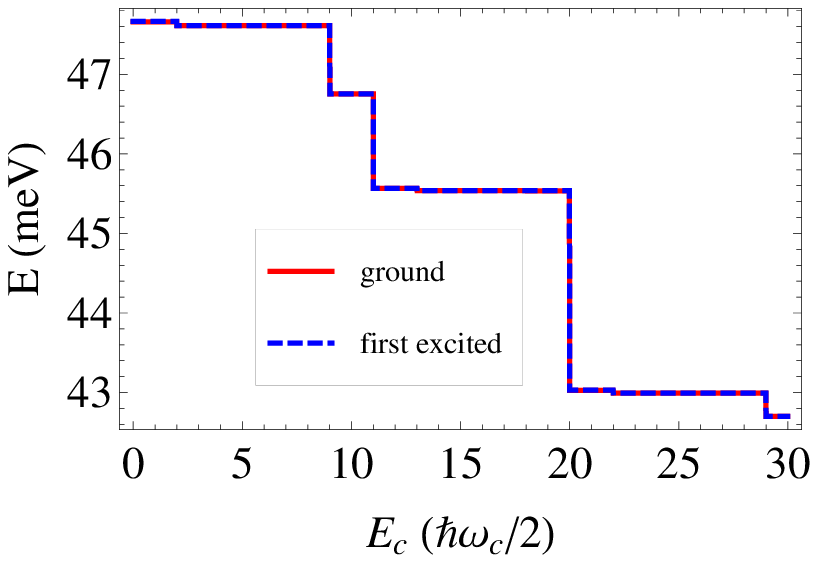}
\caption{Convergence of the lowest two energies of the CI-1 method for a 6-electron DQD using the ``medium'' (upper pane) and ``large'' (lower pane) dot parameters of Table \ref{tblParameters}  with zero tilt ($\epsilon=0$). \label{figCI1convMedLarge}}
\end{center}
\end{figure}

Next, we turn to the convergence behavior of the ``medium'' and ``large''-dot DQDs of the text, given by the 6-electron DQD parameters of Table \ref{tblParameters}.  Figure \ref{figCI1convMedLarge} shows the convergence of the CI-1 ground state at zero tilt.  It is difficult to conclude that convergence is obtained from these plots, as possibility of small steps beyond $E_c=30\hbar\omega_c/2$ cannot be eliminated, and we prefer the CI-2 for these dot sizes because of its smoother convergence and lower ground state energy. 

\begin{figure}[h]
\begin{center}
\includegraphics[width=1.5in,angle=270]{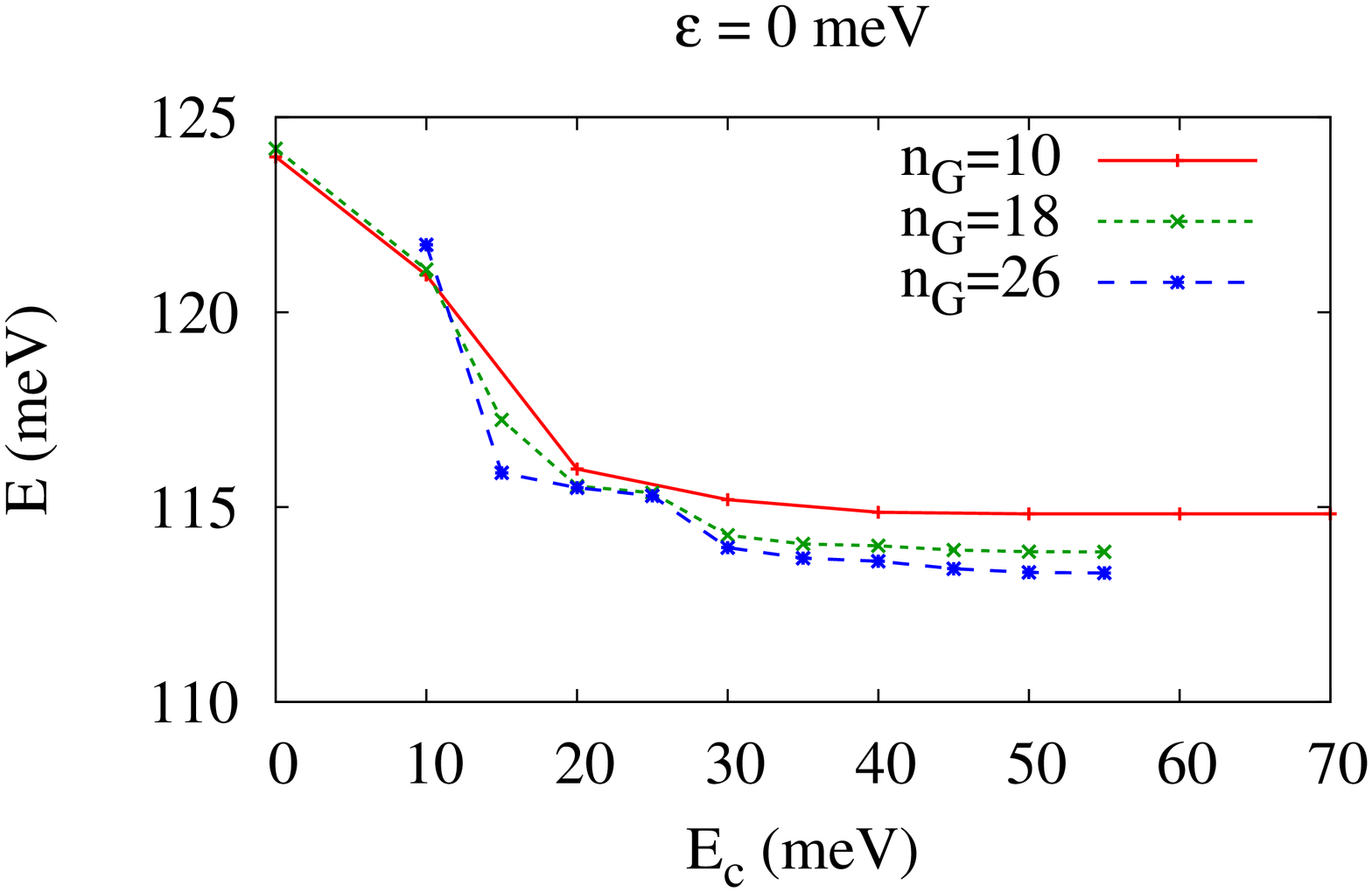}
\includegraphics[width=1.5in,angle=270]{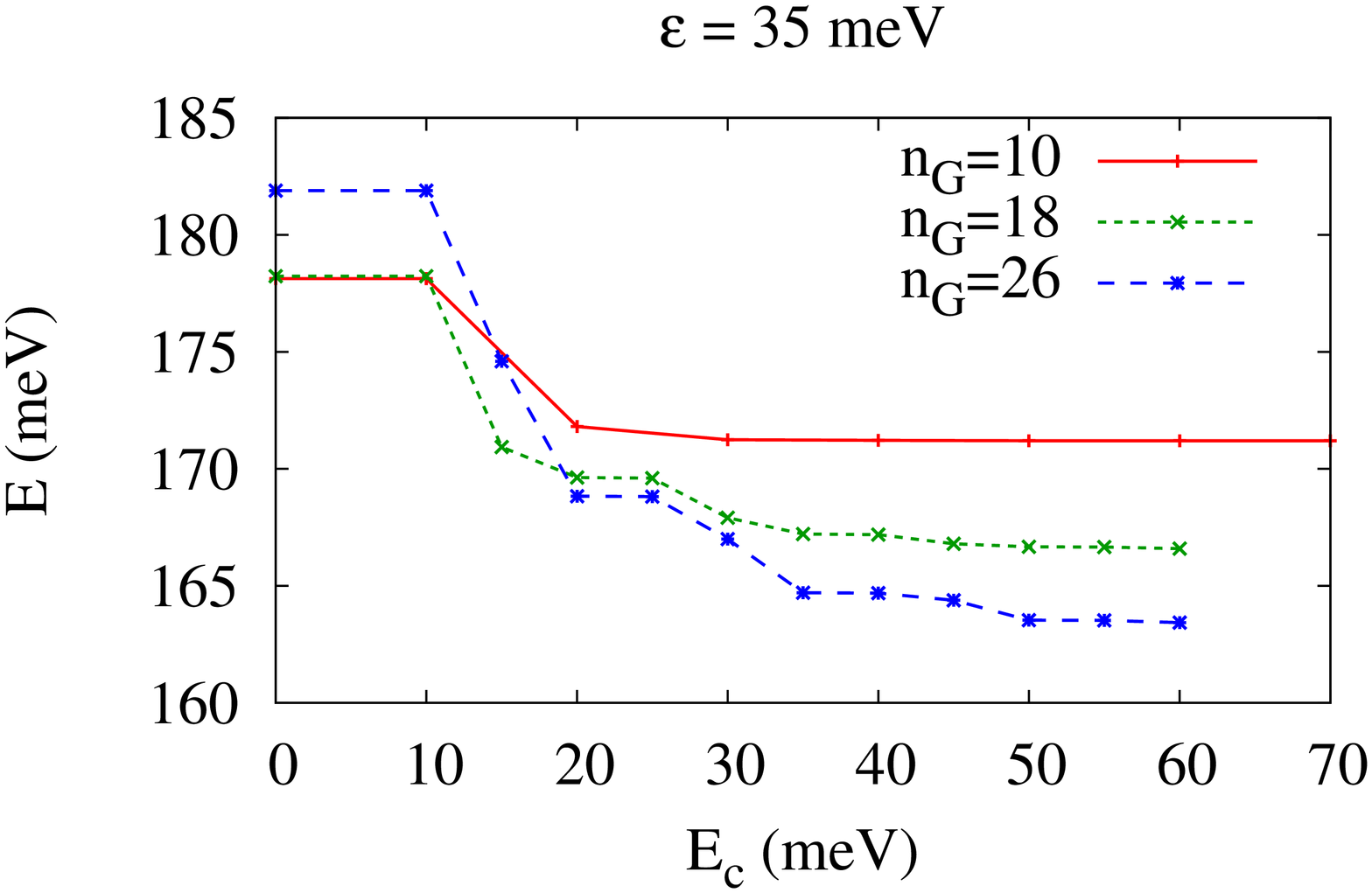}
\caption{Convergence of the lowest singlet energy (S0) of the CI-2 method for a 6-electron DQD in the (1,1) regime (upper pane) and (0,6) regime (lower pane).  Medium dot parameters $E_0=7.28\meV$, $L=35\nm$, and $B=847\mT$, are used and the tilt $\epsilon$ is indicated above each plot.\label{figCI2convMedium}}
\end{center}
\end{figure}

\begin{figure}[h]
\begin{center}
\includegraphics[width=1.5in,angle=270]{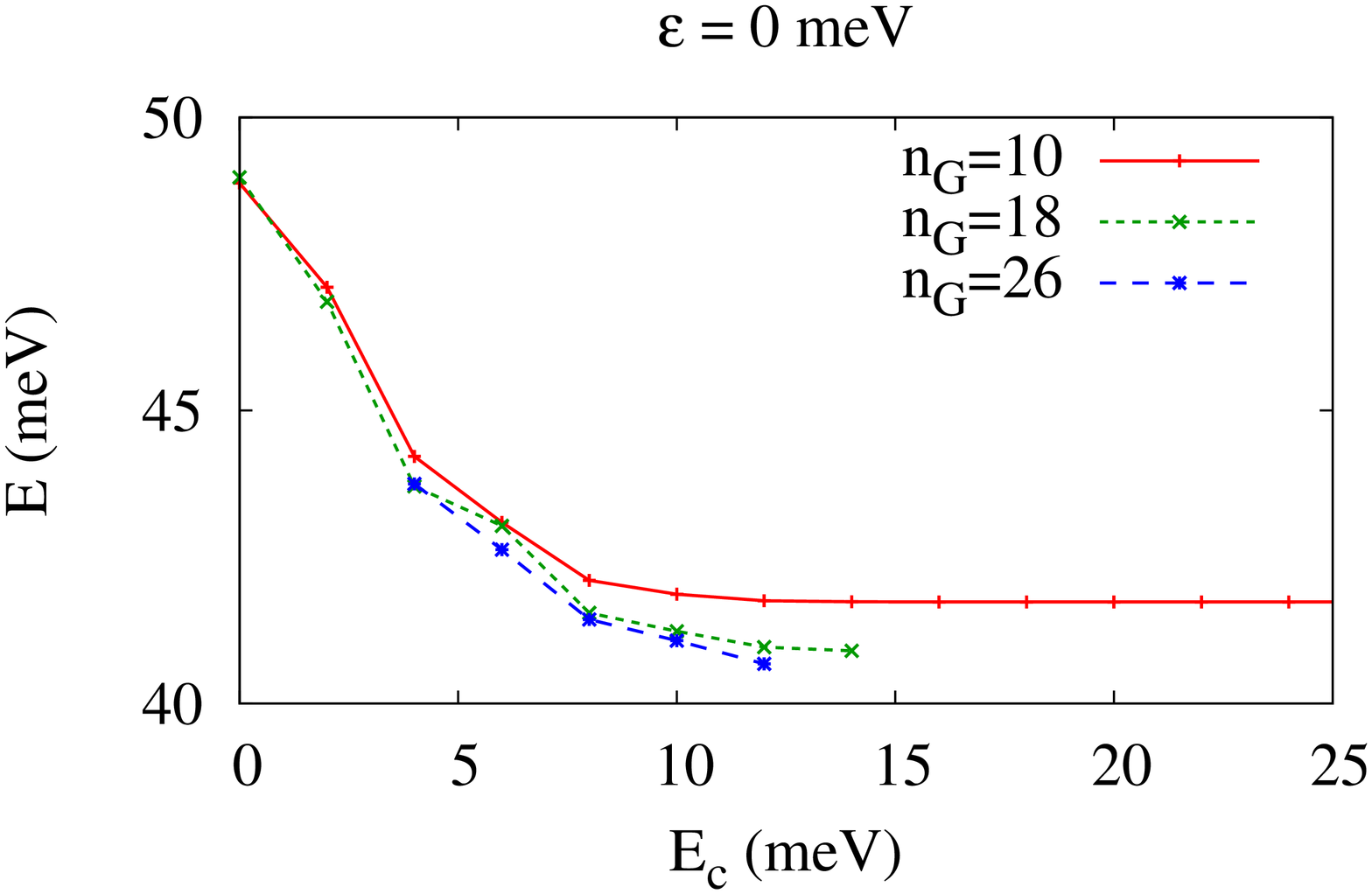}
\includegraphics[width=1.5in,angle=270]{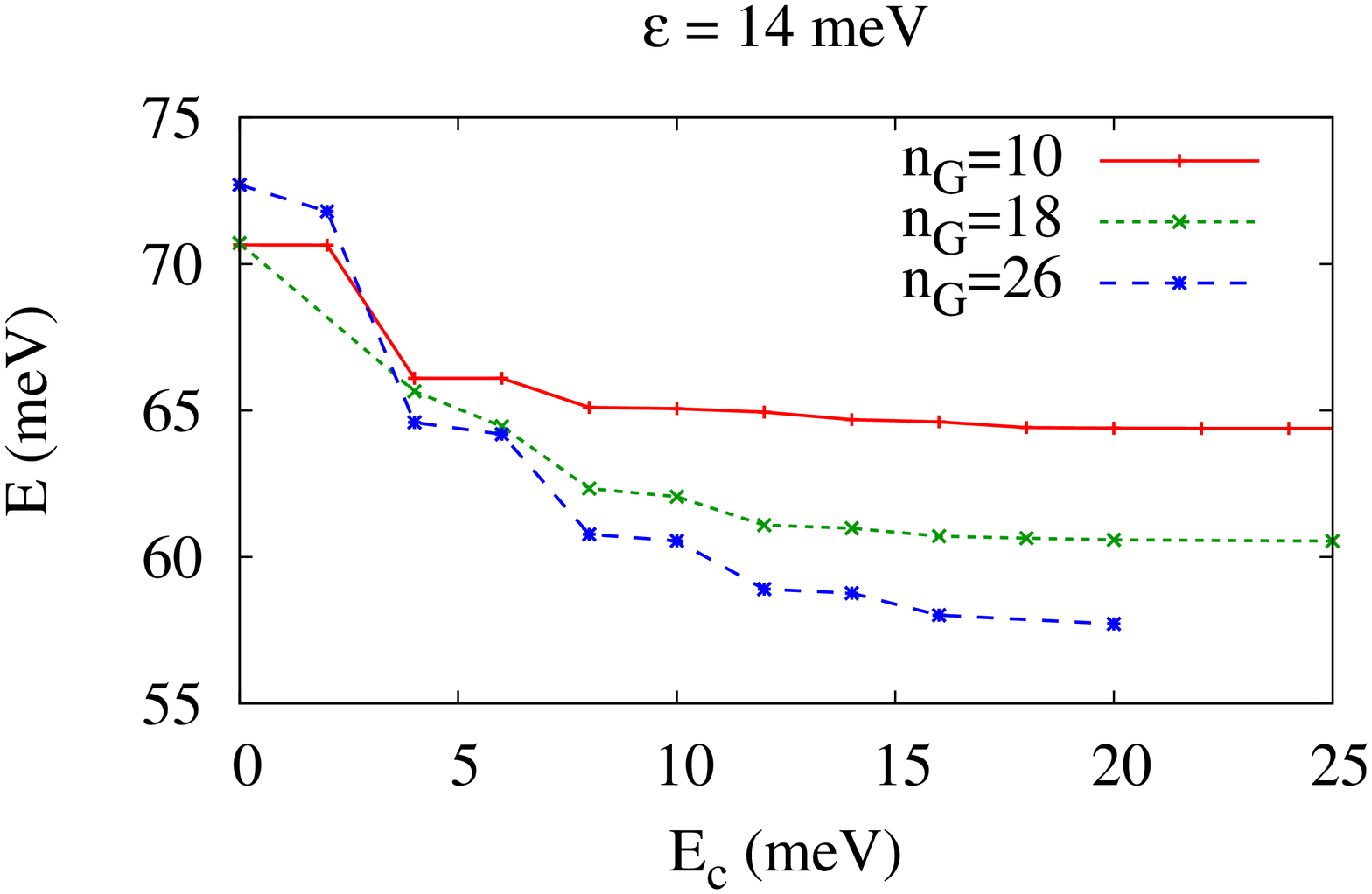}
\caption{Convergence of the lowest singlet energy (S0) of the CI-2 method for a 6-electron DQD in the (1,1) regime (upper pane) and (0,6) regime (lower pane).  Large dot parameters $E_0=1.82\meV$, $L=60\nm$, and $B=212\mT$, are used and the tilt $\epsilon$ is indicated above each plot.\label{figCI2convLarge}}
\end{center}
\end{figure}

 Figures \ref{figCI2convMedium} and \ref{figCI2convLarge} show the convergence behavior of CI-2 in the (3,3) and (0,6) regions of detuning similar to Fig.~\ref{figCI2convSmall}.  We notice progressively less step-like jumps in the energy curves as the dots become larger, indicating that there is the shell-structure picture is becoming less valid and relevant single particle levels are more numerous and spread out over energy.  We again note the larger differences in converged energy in the (0,6) regime, and that in the large-dot DQD case the energy if $n_G=18$ is barely just barely converging with respect to $E_c$.  This motivates the care taken in the text in the interpretation of the large dot results.

\section{Comparison of Fock-Darwin and $s$-type Gaussian CI bases\label{CIAppendix}}

\begin{figure}[h]
\begin{center}
\includegraphics[width=2in,angle=270]{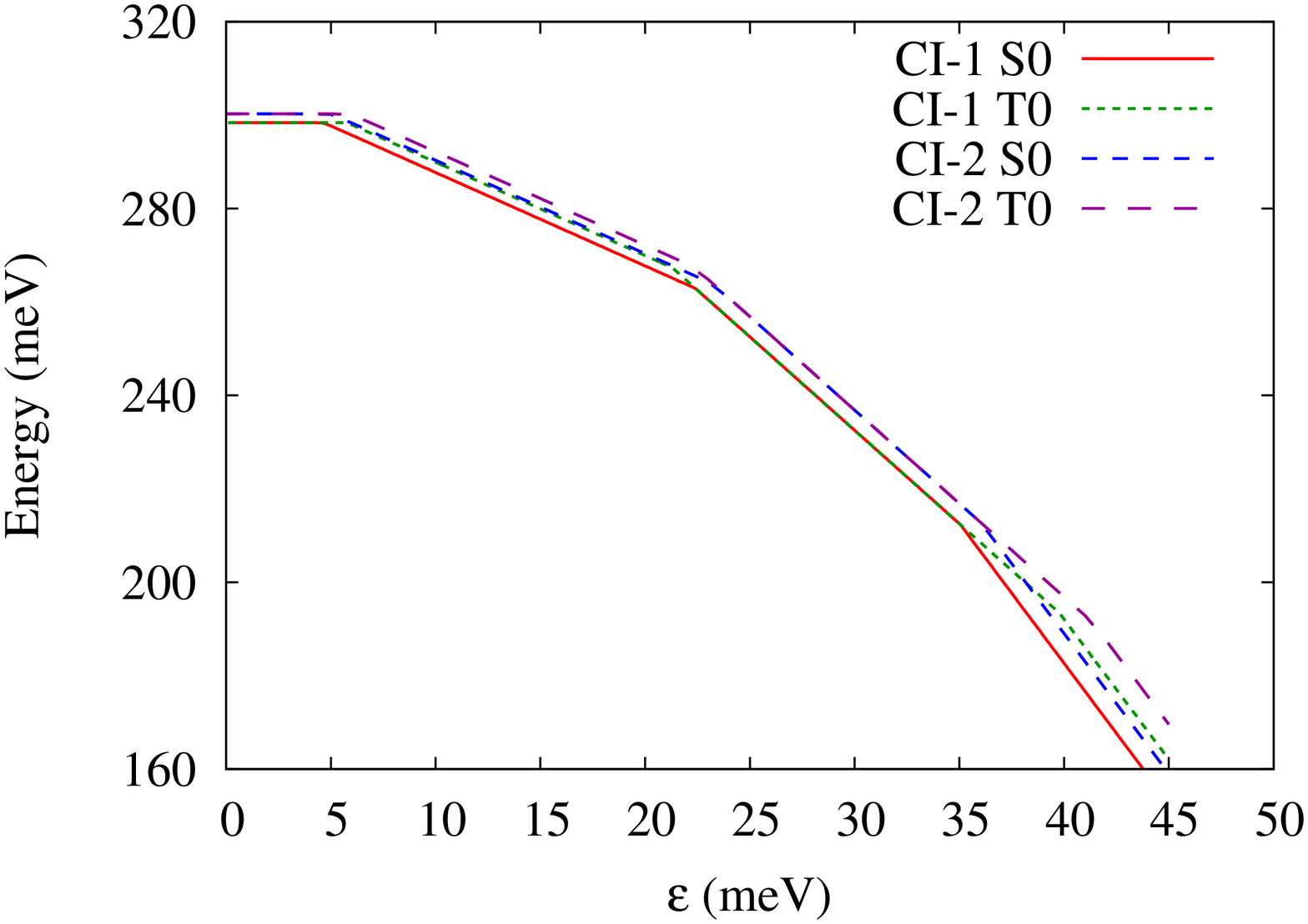}
\includegraphics[width=2in,angle=270]{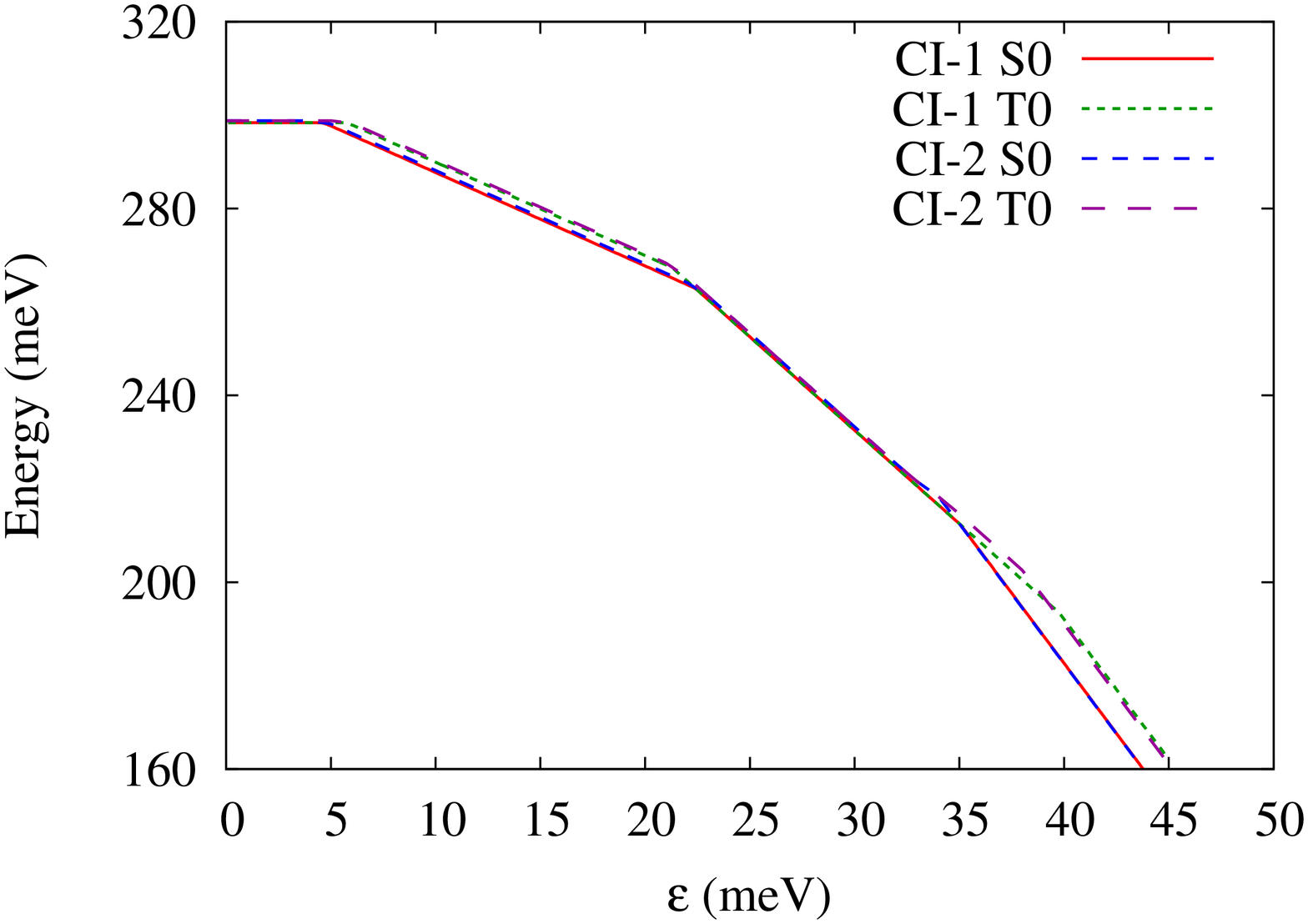}
\caption{Comparison of 6 electron energies produced by the two different CI implementations, CI-1 and CI-2, when CI-2 uses $n_G=10$ $s$-type Gaussians in the upper plot and $n_G=26$ in the lower plot.  Small-dot parameters ($E_0=24.26\meV$, $L=20\nm$, $B=2.8\Tesla$) are used.\label{figCIsComparison}}
\end{center}
\end{figure}

In this appendix we compare the two configuration interaction implementations, CI-1 and CI-2, used in this work.  While not of direct relevance to the results other than giving further confirmation and corroboration of each implementations correctness, we feel a detailed comparison will still be of interest in its own right.  Comparisons are made using the ``small-dot'' DQD parameters of the text, since for these parameters both CI-1, which uses Fock-Darwin states as a single-particle basis, and CI-2, which uses a array of $s$-type Gaussian functions as a basis, converge well with cutoff energy $E_c$ and the number of basis elements.
Figure \ref{figCIsComparison} compares the energy of the lowest singlet and triplet states (S0 and T0) when CI-2 uses 10 Gaussians and 26 Gaussians.  The agreement is worst in the large tilt region, and CI-1 finds a lower ground state energy by a few milli-electron volts in this regime in the 10-Gaussian case.  In the 26-Gaussian case the agreement is good to a small fraction of a milli-electron volt throughout the detuning range.  The 26-Gaussian CI-2, however, is much more expensive than the CI-1 in terms of the number of configurations considered and total computation time.  This highlights the fact that the CI-1 is better suited to small-dot DQDs as discussed in the text.

For larger dot sizes, such as the medium and large dot parameters in the text, we find that the CI-2 method produces lower ground state energies than CI-1.  We believe this is due to the fact that products of Fock-Darwin orbitals are less good variational approximations of the true multi-electron wave functions in larger dots, whereas the single particle basis used by CI-2 is more generically appropriate.


Overall, the agreement between the two independently developed CI codes gives us confidence in the correctness of the implementations and lends credence to the accuracy of a converged result using either implementation.

\bibliography{shellDQDqubit}

\end{document}